\documentclass[twocolumn]{aastex631}
\usepackage[caption=false]{subfig}
\usepackage{rotating}

\newcommand\HI{H\textsc{i}}
\newcommand\draftone[1]{\textcolor{black}{#1}}
\newcommand\drafttwo[1]{\textcolor{black}{#1}}
\newcommand\draftthree[1]{\textcolor{black}{#1}}
\newcommand\refrep[1]{\textcolor{black}{#1}}
\newcommand\draftfour[1]{\textcolor{black}{#1}}

\shorttitle{\drafttwo{VERTICO \textsc{II}: effects of \HI-identified environmental mechanisms on molecular gas}}
\shortauthors{Zabel et al.}
\graphicspath{{./}{figures/}}

\begin{document}

\title{\draftthree{VERTICO \textsc{II}: How H\textnormal{\textsc{i}}-identified environmental mechanisms affect the molecular gas in cluster galaxies}}

\correspondingauthor{Nikki Zabel}
\email{zabel@astro.rug.nl}

\author[0000-0001-7732-5338]{Nikki Zabel}
\affiliation{Kapteyn Astronomical Institute, University of Groningen PO Box 800, NL-9700 AV Groningen, The Netherlands}

\author[0000-0003-1845-0934]{Toby Brown}\affiliation{Herzberg Astronomy and Astrophysics Research Centre, National Research Council of Canada, 5071 West Saanich Rd, Victoria, BC, V9E 2E7, Canada}\affiliation{Department of Physics \& Astronomy, McMaster University, 1280 Main Street W, Hamilton, ON, L8S 4M1, Canada}

\author[0000-0001-5817-0991]{Christine D. Wilson}\affiliation{Department of Physics \& Astronomy, McMaster University, 1280 Main Street W, Hamilton, ON, L8S 4M1, Canada}

\author{Timothy A. Davis}\affiliation{School of Physics \& Astronomy, Cardiff University, Queens Buildings, The Parade, Cardiff CF24 3AA, UK}

\author{Luca Cortese}\affiliation{International Centre for Radio Astronomy Research, The University of Western Australia, 35 Stirling Hwy, 6009 Crawley, WA, Australia }\affiliation{ARC Centre of Excellence for All Sky Astrophysics in 3 Dimensions (ASTRO 3D), Australia}

\author{Laura C. Parker}\affiliation{Department of Physics \& Astronomy, McMaster University, 1280 Main Street W, Hamilton, ON, L8S 4M1, Canada}

\author{Alessandro Boselli}\affiliation{Aix-Marseille Universit\'{e}, CNRS, CNES, LAM, Marseille, France}

\author{Barbara Catinella}\affiliation{International Centre for Radio Astronomy Research, The University of Western Australia, 35 Stirling Hwy, 6009 Crawley, WA, Australia }\affiliation{ARC Centre of Excellence for All Sky Astrophysics in 3 Dimensions (ASTRO 3D), Australia}

\author{Ryan Chown}\affiliation{Department of Physics \& Astronomy, McMaster University, 1280 Main Street W, Hamilton, ON, L8S 4M1, Canada}

\author[0000-0003-1440-8552]{Aeree Chung}
\affiliation{Department of Astronomy, Yonsei University, 50 Yonsei-ro, Seodaemun-gu, Seoul, 03722, South Korea}

\author{Tirna Deb}
\affiliation{Kapteyn Astronomical Institute, University of Groningen PO Box 800, NL-9700 AV Groningen, The Netherlands}

\author{Sara L. Ellison}
\affiliation{Department of Physics \& Astronomy, University of Victoria, Finnerty Road, Victoria, BC, V8P 1A1, Canada}

\author{Mar\'ia J. Jim\'enez-Donaire}
\affiliation{Observatorio Astronómico Nacional (IGN), C/Alfonso XII, 3, E-28014 Madrid, Spain}
\affiliation{Centro de Desarrollos Tecnológicos, Observatorio de Yebes (IGN), 19141 Yebes, Guadalajara, Spain}

\author{Bumhyun Lee}
\affiliation{\draftfour{Korea Astronomy and Space Science Institute, 776 Daedeokdae-ro, Yuseong-gu, Daejeon 34055, Republic of Korea}}
\affiliation{\draftfour{Kavli Institute for Astronomy and Astrophysics, Peking University, Beijing 100871, China}}

\author{Ian D. Roberts}
\affiliation{Leiden Observatory, Leiden University, PO Box 9513, 2300 RA Leiden, The Netherlands}

\author{Kristine Spekkens}
\affiliation{Royal Military College of Canada, PO Box 17000, Station Forces, Kingston, ON, Canada K7K 7B4}

\author{Adam R.H. Stevens}
\affiliation{International Centre for Radio Astronomy Research, The University of Western Australia, 35 Stirling Hwy, 6009 Crawley, WA, Australia}

\author[0000-0003-0080-8547]{Mallory Thorp}
\affiliation{Department of Physics \& Astronomy, University of Victoria, Finnerty Road, Victoria, BC, V8P 1A1, Canada}

\author{Stephanie Tonnesen}
\affiliation{CCA, Flatiron Institute, 162 5th Ave, New York NY 10010}

\author[0000-0002-5877-379X]{Vicente Villanueva}
\affiliation{Department of Astronomy, University of Maryland, College Park, MD 20742, USA}



\begin{abstract}
In this VERTICO early science paper we explore in detail how environmental mechanisms\draftfour{,} identified in \HI\draftfour{,} affect the resolved properties of molecular gas reservoirs in cluster galaxies. The molecular gas is probed using ALMA ACA (+TP) observations of $^{12}$CO(2\draftthree{--}1) in 51 spiral galaxies  in the Virgo cluster \drafttwo{(of which 49 are detected)}, all of which are included in the VIVA \drafttwo{\HI}\ survey. The sample spans a stellar mass range of $9 \leq \text{log}\,M_\star / \text{M}_\odot \leq 11$. We study molecular gas radial profiles, isodensity radii, and surface densities as a function of galaxy \HI\ deficiency and morphology. \drafttwo{There is \draftthree{a weak} correlation between global \HI\ and H$_2$ deficiencies,} and resolved properties of molecular gas correlate with \HI\ deficiency\draftfour{:} galaxies that have large \HI\ deficiencies have relatively steep and truncated molecular gas radial profiles, which is due to the removal of low-surface density molecular gas \draftfour{on} the outskirts. Therefore, while the environmental mechanisms observed in \HI\ also affect molecular gas reservoirs, \draftthree{there is only a moderate reduction of the total amount of molecular gas.}
\end{abstract}



\section{Introduction} \label{sec:intro}
Hosting hundreds to thousands of galaxies bound by a dark matter halo, galaxy clusters form the largest \drafttwo{virialised} structures in the Universe. In these high-density environments\drafttwo{,} galaxies evolve differently from their isolated counterparts\drafttwo{, with} galaxy clusters in the local Universe typically \draftthree{harbouring} a much higher fraction of passive galaxies than the field population \citep[e.g.][]{Oemler1974, Dressler1980, Goto2003}. This implies that \draftthree{some} galaxies residing in clusters \drafttwo{slow down (``quench'')} their star formation relatively quickly. Various environmental processes are known to be able to contribute to this premature quenching, such as ram pressure stripping (RPS, \citealt{Gunn1972}, \draftthree{see also \citealt{Cortese2021} and \citealt{Boselli2021} for recent reviews}), \draftthree{the lack of fresh gas available for accretion, e.g. from the circum-galactic medium, causing a galaxy to deplete its gas reservoir} \citep[also referred to as ``starvation'',][]{Larson1980}, galaxy-galaxy interactions \citep{Moore1996}, tidal interactions, and thermal evaporation and viscous stripping (\citealt{Cowie1977, Nulsen1982}, \draftthree{see also \citealt{Cortese2021} for extensive descriptions of each environmental mechanism)}. The relative importance of these mechanisms, and to which extent this varies between clusters, is not yet well-understood. 

Typically distributed within an extended disc ($D_{\text{\HI}} \sim 1-2\ D_{25}$, e.g. \citealt{Verheijen2001}) and relatively loosely bound, atomic gas (\HI) is \drafttwo{susceptible to} environmental processes. Cluster galaxies are often \HI-deficient (i.e. they contain less \HI\ than expected from their optical size), and have truncated \draftfour{and/}or asymmetric \HI\ discs \citep[e.g.][]{Haynes1984, Giovanelli1985, Cayatte1990, Solanes2001, Schroder2001, Waugh2002, Gavazzi2005, Rasmussen2006, Hughes2009, Chung2009, Odekon2016, Yoon2017, Loni2021}. \draftfour{Since atomic gas provides the fuel for molecular clouds, which will eventually collapse into stars, this removal of \HI\ might lead to the quenching of star formation.}

More recently, it has been shown that, \drafttwo{despite its more tightly bound and centrally located nature (e.g. \citealt{Davis2013})}, molecular gas can also be \emph{directly} affected by environmental processes. \drafttwo{For example, molecular gas discs in cluster galaxies have been found to be truncated and/or morphologically and kinematically disturbed, while molecular gas fractions have been found to be deficient compared to field galaxies at fixed stellar mass} \citep[e.g.][]{Vollmer2008, Fumagalli2009, Boselli2014, Lee2017, Lee2018, Zabel2019, Cramer2021, Brown2021}. \draftfour{Additionally, molecular gas fractions can also be \emph{enhanced} in cluster galaxies, as a result of ram pressure, which can facilitate the conversion of atomic into molecular gas (e.g. \citealt{Moretti2020})}. 

It is not yet clear to what extent the processes acting on the atomic gas also affect the molecular gas and, if they do, on what relative timescales. It is possible that\drafttwo{:} a) atomic and molecular gas are removed and disturbed simultaneously by the same environmental processes \drafttwo{(for example by galaxy-galaxy interactions)}, b) the majority of the atomic gas is removed before the molecular gas \drafttwo{(for example by RPS)}, or c) atomic and molecular gas are affected by different environmental processes entirely \draftthree{(for example, molecular gas is depleted through starvation while atomic gas is being stripped)}. Which of these options is closest to reality has strong implications for the process of environmental quenching, and the timescales on which this takes place.

Whether the molecular gas reservoirs of Virgo galaxies are different from those of their counterparts in the field has been a matter of debate \citep{Stark1986, Kenney1989, Boselli1995, Mok2016, Mok2017}. \citet{Boselli2014} studied the H$_2$ content of \drafttwo{$\sim$75 spiral galaxies in} the Virgo cluster using the \textit{Herschel} Reference Survey (HRS, \citealt{Boselli2010}) and compared them to the remaining field galaxies in the HRS. They \drafttwo{find} that, on average, \HI-deficient Virgo cluster galaxies have less molecular gas than \HI-normal field galaxies. Moreover, they find a weak but statistically significant increase in H$_2$ deficiency as a function of \HI\ deficiency. This suggests that both gas phases are affected simultaneously \drafttwo{in the Virgo cluster}. Resolved observations of 17 of these HRS Virgo galaxies \citep{Kuno2007} show a decrease in the extent of the molecular gas disc with \drafttwo{increasing} \HI\ deficiency, which implies outside-in stripping of the molecular gas. On the other hand, \citet{Loni2021} \draftthree{find} significant scatter in H$_2$/\HI\ \draftfour{mass} ratios in galaxies in the Fornax cluster (including both upper and lower limits), suggesting that the relative effects of environmental mechanisms on both gas phases are not straightforward. 

This work is one of the early science papers of the Atacama Large Millimeter/submillimeter Array (ALMA) large programme the ``Virgo Environment Traced in CO'' (VERTICO, \citealt{Brown2021}, hereafter referred to as B21). VERTICO comprises \drafttwo{homogeneous} CO(2-1) observations of 51 spiral galaxies in the Virgo cluster (of which 49 are detected), and \draftthree{is} designed to systematically study the physical mechanisms that drive galaxy evolution in dense environments. Because the VERTICO sample is \drafttwo{selected from} the Very Large Array Imaging of Virgo in Atomic gas survey (VIVA, \citealt{Chung2009}, see \S \ref{sec:sample}), ancillary \HI\ imaging is available for all CO\drafttwo{-}detected galaxies in the sample. In this work, we study the resolved atomic and molecular gas in the 49 Virgo galaxies detected in CO with VERTICO.

The Virgo cluster is a young and dynamically active cluster, with several sub-structures and infalling groups \citep[\draftthree{e.g.}][]{Gavazzi1999, Solanes2002, Boselli2014c, Lisker2018}. \draftthree{As the closest galaxy cluster} to us (at $\sim$16.5 Mpc, \citealt{Mei2007}), it has been studied extensively in a range of wavelengths \citep[see also the references in B21]{Binggeli1985, McLaughlin1999, Fouque2001, Davies2010, Ferrarese2012, Mihos2017, Boselli2011, Boselli2018}. Following B21, \draftthree{we adopt} a common distance of 16.5 Mpc for all galaxies in the sample.

In this work, we study how molecular gas is affected by environment in galaxies in the Virgo cluster. We make use of the homogeneity and resolution of VERTICO to study resolved properties of the molecular gas, such as \draftthree{sizes and radial profiles}, and compare them to those of \draftthree{control} galaxies. \draftthree{We probe the extent to which a galaxy is affected by environment, and \draftthree{the mechanism by which it is affected} (if it is possible to identify this), using \HI\ \draftfour{data} from the VIVA survey \citep{Chung2009, Yoon2017}.}

This paper is organised as follows. In \S \ref{sec:sample} the sample, observations, and data reduction are described. \draftthree{The control} samples used to compare our results to are introduced here. \S \ref{sec:methods} contains the methods used, as well as a description of various definitions adopted throughout this work. \drafttwo{Deficiency parameters are defined for both the atomic and molecular gas, and the calculation of radial profiles, H$_2$ radii, and median H$_2$ surface densities is \draftthree{also described here}.} In \S \ref{sec:results} we describe the results, and provide brief interpretations, while in \S \ref{sec:discussion} we provide a more thorough interpretation and discussion, \drafttwo{including comparison to previous work}. Finally, a short summary, as well as an itemised list of the findings from this work, are given in \S \ref{sec:summary}.

\section{Sample, observations \& data reduction} \label{sec:sample}
\label{sec:sample}
\draftthree{The sample used in this work consists of the 49 CO-detected VERTICO galaxies.} The sample selection and data reduction of VERTICO are described in detail in B21, and are summarised below. 

VERTICO targeted the CO(2-1) line in 51 late-type galaxies that were part of the VIVA survey \citep{Chung2009} using the ALMA Atacama Compact Array (ACA). Of these, 15 galaxies that were already observed with the ACA were taken from the ALMA archive. Archival galaxies primarily come from the ALMA component of the Physics at High Angular resolution in Nearby Galaxies project (PHANGS-ALMA; \citealt{Leroy2021a}, various project IDs; see table 2 in that work). One galaxy, NGC4402, comes from \citet[][project ID 2016.1.00912.S]{Cramer2020}. The final sample includes galaxies undergoing a variety of environmental effects, as identified in their \HI\ imaging \drafttwo{\citep{Chung2009}}. \drafttwo{Our sample} spans a stellar mass range of $10^{\draftthree{9}} \leq M_\star/\text{M}_\odot \leq 10^{11}$. \drafttwo{\draftthree{B21} calculated inclination and position angles} from fits to Sloan Digital Sky Survey (SDSS, \citealt{York2000, Alam2015}) \textit{r}-band images \drafttwo{(see table 1 in B21)}.

The 36 galaxies that \draftthree{were} not yet on the ALMA archive were observed during Cycle 7 under program ID 2019.1.00763.L. Total Power (TP) observations were added for 25 targets for which the CO was expected to extend beyond the largest recoverable angular size of 29$^{\prime\prime}$. 

\draftone{For the 36 Cycle-7 galaxies we used the calibrated \emph{uv}-data as delivered by ALMA. The raw \emph{uv}-data of the remaining 15 objects were recalibrated using the Common Astronomy Software Applications package (CASA, \citealt{McMullin2007}), \drafttwo{using the appropriate version}. For the imaging of both the ACA and the TP data the PHANGS-ALMA Imaging Pipeline Version 1.0 and the PHANGS-ALMA TP Pipeline were used, respectively \citep{Herrera2020, Leroy2021b}. Three modifications were implemented in the Imaging Pipeline to optimise for the VERTICO sample, which are described in detail in \S 3.1 of B21. A Briggs weighting scheme was adopted \citep{Briggs1995}, with a robust parameter of 0.5. The total-power data were processed at their native velocity resolution of $\sim 3$~km~s$^{-1}$, and no other modifications were made to the TP Pipeline. For the PHANGS-VERTICO galaxies, calibrated TP cubes were kindly provided to us by Adam Leroy on behalf of the PHANGS-ALMA team in private communication. Finally, the total-power data were combined with the ACA data via feathering using the PHANGS-ALMA pipeline \citep{Leroy2021b}.}

The median spatial resolution of the resulting data cubes is $\sim$ 8$^{\prime\prime}$ ($\sim$640 pc at the distance of Virgo), and the final velocity resolution is 10~km~s$^{-1}$. The flux calibration uncertainty is typically $5-10\%$\footnote{\draftfour{See the ALMA Technical Handbook: \url{https://almascience.nrao.edu/documents-and-tools/cycle8/alma-technical-handbook}}}. The typical \drafttwo{root-mean-square (rms) sensitivity} reached with the ACA, after \draftone{an average integration time of 3.7 hours on source,} is 10.6 mJy beam$^{-1}$ per $10 ~ {\rm km ~ s^{-1}}$ channel. All galaxies were detected in CO except IC3418 and VCC1581, for which we calculated 3$\sigma$ upper limits. \drafttwo{Although the presence of \HI\ emission \draftthree{in IC3418} was confirmed later by \citet{Kenney2014}, its image is not available (unlike VCC1581) and it is not considered in this work beyond Figure \ref{fig:H2_frac}.}

\subsection{\draftthree{Control} samples} \label{sub:field_samples}
We use two types of comparison samples. \draftfour{First, the} \draftthree{extended} Galaxy Evolution Explorer (GALEX) Arecibo SDSS Survey (xGASS; \citealt{Catinella2018}) and the extended CO Legacy Database for \textit{GALEX} Arecibo SDSS Survey \citep[xCOLD GASS;][]{Saintonge2017} are used to compare global \HI\ and H$_2$ masses \draftthree{in VERTICO} to those of \draftthree{the general population of} galaxies at fixed stellar mass, and estimate \HI\ and H$_2$ deficiencies. After applying a small number of selection criteria \drafttwo{(described below)}, these samples comprise \draftfour{several} hundreds of galaxies, thus providing a good representation of the galaxy population throughout the relevant stellar mass range. \draftfour{Second, for} a comparison of \emph{resolved} properties of VERTICO galaxies with \draftthree{control} galaxies\drafttwo{,} we use the \draftthree{HEterodyne Receiver Array} CO-Line Extragalactic Survey (HERACLES, \citealt{Leroy2009}). This sample is described in detail in \S \ref{sub:heracles}. 

\drafttwo{In Figure \ref{fig:SFMS} all three control samples, as well as the VERTICO sample, are shown on the SFR\draftthree{--}$M_\star$ plane. \draftfour{Stellar masses and star formation rates for x(COLD) GASS were taken from GALEX-SDSS-WISE Legacy Catalog (GSWLC, \citealt{Salim2016}), as these are consistent with those from the $z=0$ Multi-wavelength Galaxy Synthesis ($z$0MGS, \citealt{Leroy2019}), which are adopted for the VERTICO sample (see \S \ref{subsub:stellar_masses}). Galaxies for which no 22 $\mu$m detection is available (used for deriving SFRs) have upper limits on their SFRs are indicated with down-pointing triangles.} VERTICO galaxies occupy a similar region in this plane as star forming (i.e. non-quiescent) x(COLD) GASS galaxies. HERACLES galaxies have enhanced SFRs at fixed stellar mass, which is likely due to selection effects (\S \ref{sub:heracles})} \draftthree{This implies that HERACLES galaxies may not be typical field galaxies, and may have increased gas fractions, if they follow the Kennicutt-Schmidt relation \citep{Schmidt1959, Kennicutt1998}. We \draftfour{should} keep this in mind when we compare results from the VERTICO sample to those from the HERACLES sample.}

\subsubsection{\drafttwo{xGASS \& xCOLD GASS}} \label{subsub:x/coldgass}
\refrep{As described in \S \ref{sec:sample}, the VERTICO sample consists exclusively of late-type galaxies. In the field, such galaxies are expected to lie on, or close to, the star formation main sequence (SFMS). Therefore, to ensure a fair comparison with VERTICO, we only consider galaxies from xGASS and xCOLD GASS within 2$\sigma$ from the SFMS \citep{Elbaz2007}.} \draftfour{Furthermore}, galaxies with confused \HI\ emission (i.e. the \HI\ emission from multiple sources cannot be separated reliably), as identified by \draftfour{\citet{Catinella2018}}, are eliminated from the xGASS sample. After applying these selection criteria, the comparison sample from xGASS consists of \draftthree{541} galaxies with detected \HI, and \draftthree{33 with} upper limits.

Molecular gas masses from xCOLD GASS were corrected for the difference in $\alpha_{\text{CO}}$ \draftthree{(\draftfour{see \S \ref{sub:H2_deficiency};} xCOLD GASS uses the metallicity-dependent $\alpha_{\text{CO}}$ from \citealt{Accurso2017}).} The final comparison sample consists of 303 galaxies, \draftthree{44 of which are upper limits.}

\drafttwo{Upper limits are taken into account in the calculation of the median H$_2$ mass using survival analysis (a statistical method designed to take into account ``censored'' data: measurements for which only an upper or lower limit is available). \drafttwo{We used the Kaplan\draftthree{--}Meier estimator, implemented in the \texttt{Python} package \texttt{Lifelines} \citep{Davidson-Pilon2019}, to estimate the true cumulative distribution of H$_2$ mass with upper limits taken into account. The median H$_2$ mass was estimated from the resulting distribution.} The Kaplan\draftthree{--}Meier estimator takes censored data into account assuming they follow a similar distribution to the measured data. Finally, we calculate the rolling median of the H$_2$ mass by dividing the sample into 10 stellar mass bins and using a shift of half a bin.}

\subsubsection{HERACLES: a resolved control sample} \label{sub:heracles}
\drafttwo{We compare the resolved properties of VERTICO galaxies to those from HERACLES.} \draftthree{We first remove galaxies that are also in VERTICO from the HERACLES sample (NGC4254, NGC4321, NGC4536, NGC4569, and NGC4579), as well as interacting galaxies (NGC2146, NGC2798, NGC3034, NGC3077, and NGC5713), except for those with a significantly less massive or gas-poor companion, such as NGC5194 (M51). The remaining sample consists} of 21 galaxies detected in CO. \drafttwo{Resolved quantities are calculated from the HERACLES moment 0 maps at native resolution (the typical beam size is b$_{\text{maj}} \sim 13 ^{\prime\prime}$), in the same way as those of VERTICO.} These maps are calculated from the 10 km s$^{-1}$ cubes\draftfour{,} using the VERTICO data products pipeline (B21, \S 4.1). NGC4214, NGC2841, and NGC4725 are omitted from the H$_2$ surface density analysis \drafttwo{due to their high inclination (see \S \ref{sub:densities})}. 

Global molecular gas mass fractions of the HERACLES sample are shown in Figure \ref{fig:H2_frac}. Here we can see that HERACLES galaxies are relatively gas-rich. The HERACLES sample was based on The \HI\ Nearby Galaxy Survey \citep[THINGS,][]{Walter2008}, which was in turn based on the \drafttwo{Space Infrared Telescope Facility (SIRTF)} Nearby Galaxies Survey \citep[SINGS,][]{Kennicutt2003}. By design, this sample has a flat far-infrared (FIR) luminosity distribution (as measured at 60$\mu$m with the Infrared Astrononical Satellite, IRAS). This means that, compared to the actual FIR luminosity distribution of galaxies in the local universe, FIR-faint galaxies are significantly under-represented (see also figure 6 in \citealt{Kennicutt2003}). Therefore, it is \draftone{likely} that \draftthree{the SINGS, THINGS, and HERACLES samples are} biased towards galaxies with high (atomic and molecular) gas masses \draftone{\citep[e.g.][]{Saintonge2018}}. This should be kept in mind throughout the remainder of this work.

\section{Methods} \label{sec:methods}
\subsection{\HI\ deficiency} \label{sub:HI_deficiency}
\draftone{For the majority of this work, }\HI\ deficiencies are adopted from \citet{Chung2009}, who use the Hubble type independent definition from \citet{Haynes1984b}:
\begin{equation}
\text{def}_{\text{HI},\: R_\star} = \left\langle \text{log } \bar{\Sigma}_{\text{HI}} \right\rangle - \text{log } \bar{\Sigma}_{\text{HI}}, 
\end{equation}
where $\bar{\Sigma}_{\text{HI}} \equiv S_{\text{HI}}/D^2_{\text{opt}}$, \drafttwo{where $S_{\text{HI}}$ is the \HI\ flux in Jy~km~s~$^{-1}$ and $D_{\text{opt}}$ \draftthree{is} the diameter of the optical disc in arcminutes.} \draftthree{$\bar{\Sigma}_{\text{HI}}$ is} the mean \HI\ surface density within the optical disc, and $\left\langle \text{log } \bar{\Sigma}_{\text{HI}} \right\rangle$ = 0.37 for all Hubble types. Uncertainties on the \HI\ deficiency reflect the difference with the \drafttwo{Hubble-type-dependent} deficiency for the same object (\draftone{in reality $\left\langle \text{log } \bar{\Sigma}_{\text{HI}} \right\rangle$ varies somewhat with Hubble type,} see \S5.5 in \citealt{Chung2009} for more details). \draftthree{A common distance of 16 Mpc was adopted to calculate these deficiencies \draftthree{in \citet{Chung2009}}, sufficiently similar to the distance of 16.5 Mpc adopted in this work for any differences in \HI\ measurements to be negligible.}

\begin{figure*}
	\centering
	\includegraphics[width=0.75\textwidth]{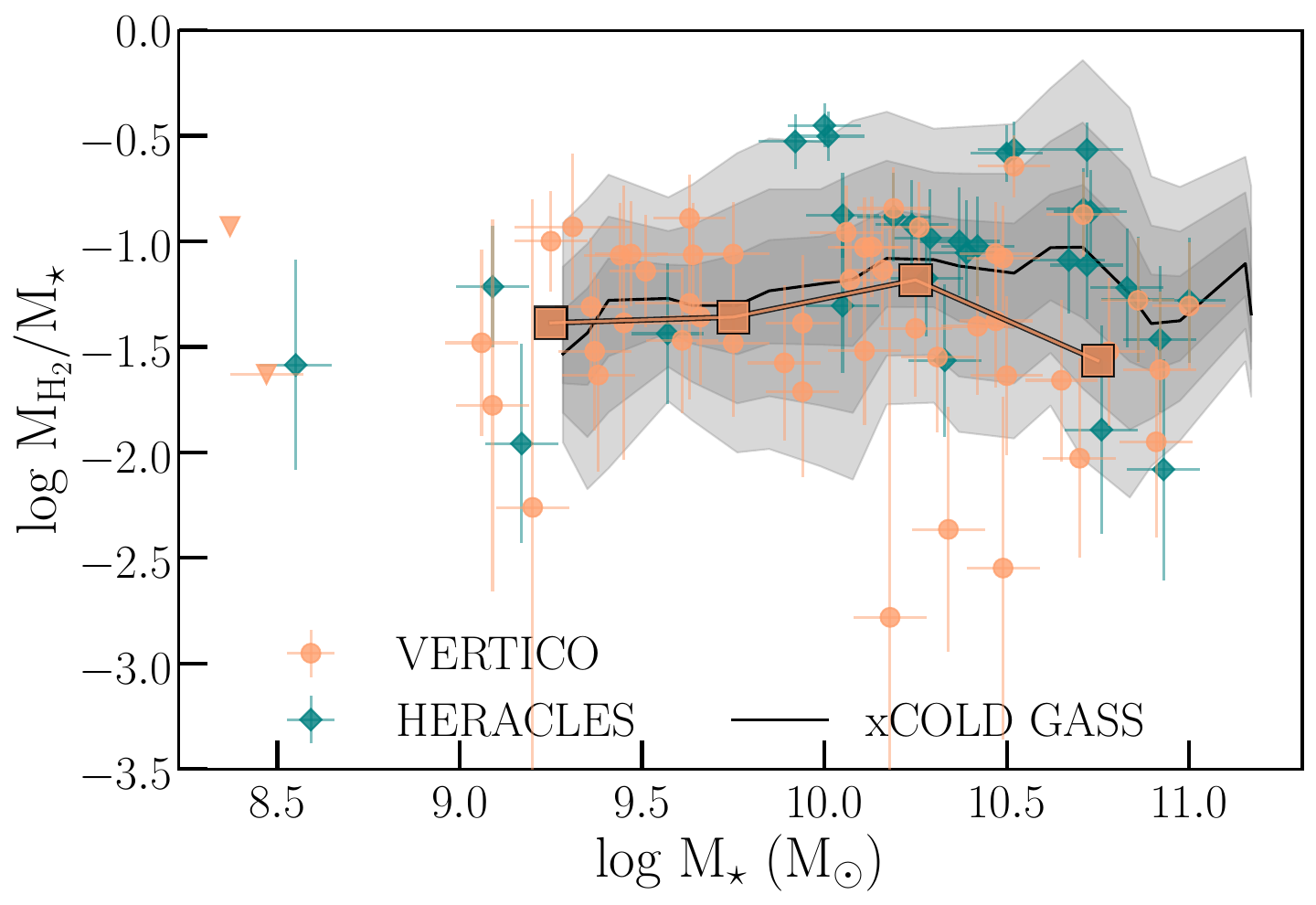}
	\caption{\draftone{H$_2$ fractions of VERTICO (orange dots) and HERACLES (teal diamonds) galaxies compared to 303 galaxies from xCOLD GASS, as a function of stellar mass. \draftthree{Downward triangles represent upper limits.} The solid black line represents the rolling median of the \drafttwo{xCOLD GASS} sample, \drafttwo{while} the shaded grey areas indicate the 1, 2, and 3 $\sigma$ spread (from dark to light, respectively). \drafttwo{The connected orange squares with black edges are the median H$_2$ fractions of VERTICO galaxies in four equal stellar mass bins \draftthree{in log space (excluding the two upper limits). VERTICO galaxies with log $M_\star \gtrsim 9.75$ log M$_\odot$ are moderately but systematically H$_2$ deficient, while low-mass VERTICO galaxies have H$_2$ fractions similar to \draftfour{or higher than} xCOLD GASS galaxies at fixed M$_\star$}.} HERACLES galaxies are relatively H$_2$-rich.}}
	\label{fig:H2_frac}
\end{figure*}

\draftone{Since such a well-defined mass\draftthree{--}size relationship does not exist for CO, the expected molecular gas mass in the equation for gas deficiency \drafttwo{(Equation \ref{eq:def_h2})} is derived from the median molecular gas mass of \draftthree{control} galaxies at fixed stellar mass, rather than the optical size of the host galaxy (see \S \ref{sub:H2_deficiency}). To ensure a fair comparison between deficiencies in both gas phases, we consider a second definition of \HI\ deficiency, similar to that for H$_2$ deficiency:}
\begin{equation}
\text{def}_{\text{\HI},\: M_\star} = \text{log } M_{\text{\HI, exp}} - \text{log } M_{\text{\HI, meas}},
\end{equation}
\draftone{where the expected \HI\ mass, $M_{\text{\HI, exp}}$, is the median \HI\ mass at fixed stellar mass from a \draftthree{control} sample from xGASS (see \ref{subsub:x/coldgass}), \draftthree{and $M_{\text{\HI, meas}}$ is the measured \HI\ mass}.}

\draftone{Uncertainties in def$_{\text{\HI},\: M_\star}$ are calculated by combining the uncertainty in the \HI\ mass and \draftfour{that} resulting from the \draftthree{uncertainty in} the stellar mass. The 1$\sigma$ spread in \HI\ mass fractions in the \draftthree{control} sample is $\sim$0.7 dex and varies somewhat with stellar mass. This spread is not taken into account in the uncertainty in $\text{def}_{\text{\HI},\: M_\star}$. The resulting \HI\ deficiencies are listed in Table \ref{tab:VERTICO-sample}.}

\subsection{H$_2$ deficiency} \label{sub:H2_deficiency}
The mass\draftthree{--}size relation for H$_2$ has not been defined as well as for \HI\ \draftone{(see e.g. B21)}. \drafttwo{\draftthree{Additionally,} the scatter in the relationship between stellar mass and molecular gas mass is smaller than that in the relationship between optical size and molecular gas mass \citep{Boselli2014}}. \draftthree{Therefore}, we use the median H$_2$ mass of a \draftthree{control} sample at fixed stellar mass to estimate the expected H$_2$ mass. \draftthree{Then the} H$_2$ deficiency can be defined as follows:
\begin{equation}
\text{def}_{\text{H}_2} = \text{log } M_{\text{H}_2, \text{exp}} - \text{log } M_{\text{H}_2, \text{meas}},
\label{eq:def_h2}
\end{equation}
where $M_{\text{H}_2, \text{exp}}$ corresponds to the expected molecular gas mass of a galaxy, and $M_{\text{H}_2, \text{meas}}$ is its measured global molecular gas mass. The \draftthree{control} sample we use to calculate $M_{\text{H}_2, \text{exp}}$ is xCOLD GASS (see \S \ref{subsub:x/coldgass}). 

Global molecular gas mass estimates for the VERTICO sample are described in detail in \S 4.4 of B21, and are summarised here. Molecular gas masses are derived from CO luminosities as follows:
\begin{equation}
M_{\text{mol}} = \frac{\alpha_{\text{CO}}}{R_{21}} L_{\text{CO}},
\label{eq:MH2}
\end{equation}
where $\alpha_{\text{CO}}$ = 4.35~M$_\odot$~pc$^{-2}$~(K~km~s$^{-1}$)$^{-1}$, corresponding to the Galactic value of $X_{\text{CO}}~=~2~\times~10^{20}$~cm$^{-2}$~(K~km~s$^{-1}$)$^{-1}$ recommended by \citet{Bolatto2013}, and ${R_{21} \equiv \text{CO}(2-1) / \text{CO}(1-0) = 0.8}$ as found by e.g. \citet{Leroy2009} \draftfour{and B21}. The CO line luminosities \drafttwo{are} calculated following \citet{Solomon2005}:
\begin{equation}
L_{\text{CO}} = 3.25 \times 10^7\ S_{\text{CO}}\ \nu^{-2}_{\text{obs}}\ D^2_L\,
\end{equation}
where $S_{\text{CO}}$ is the integrated CO line flux in Jy km s$^{-1}$, $\nu_{\text{obs}}$ the observed frequency in GHz, and $D_L$ the luminosity distance to the source in Mpc. 

Note that the adopted $\alpha_{\text{CO}}$ includes a 36\% contribution from helium. Therefore, the masses calculated here are total molecular gas masses. 

Uncertainties in H$_2$ deficiency are calculated by combining the uncertainty in the H$_2$ mass and \draftfour{that} resulting from the \draftthree{uncertainty in} the stellar mass. The 1$\sigma$ spread in molecular gas fractions in the \draftthree{control} sample is $\sim$0.4 dex and varies somewhat with stellar mass (see Figure \ref{fig:H2_frac}). This spread is not taken into account in the uncertainty in $\text{def}_{\text{H}_2}$. Molecular gas mass deficiencies are listed in Table \ref{tab:VERTICO-sample}.

Molecular gas fractions \draftthree{in the VERTICO sample are shown in Figure \ref{fig:H2_frac} as orange dots, overlaid on the sample from xCOLD GASS} (see \S \ref{subsub:x/coldgass}), whose median is shown as a solid black line, while the 1, 2, and 3 $\sigma$ spread is shown as shaded grey areas. \refrep{The medians of the VERTICO sample, in stellar mass bins of 0.5 dex, are shown as orange squares.} H$_2$ fractions of the HERACLES sample are shown as teal diamond markers. \draftthree{VERTICO galaxies with stellar masses \draftfour{log~$M_\star$/M$_\odot~\gtrsim$~9.75} are \refrep{marginally but systematically H$_2$ deficient, with the median molecular gas fractions of the stellar mass bins lying 0.1 - 0.5 dex below the xCOLD GASS rolling median.} VERTICO galaxies with stellar masses log~$M_\star$/M$_\odot~\lesssim$~9.75 have H$_2$ fractions \refrep{similar to, or marginally higher than,} those of xCOLD GASS galaxies at fixed stellar mass.} HERACLES galaxies are H$_2$-rich compared to xCOLD GASS at fixed stellar mass (see also \S \ref{sub:heracles}). 

There is a sub-sample of VERTICO galaxies that overlaps with a sub-sample of the \textit{Herschel} Reference Survey (HRS, \citealt{Boselli2010}) for which high-quality CO data \draftthree{are} available \citep{Kuno2007}. H$_2$ deficiencies for these galaxies were estimated by \citet{Boselli2014} \draftthree{similarly to this work, but where $M_{\text{H}_2, \text{exp}}$ in Equation \ref{eq:def_h2} is defined as follows:}
\begin{equation}
\text{log}\ M(H_2) = c \times \text{log}(Variable) + d,
\end{equation}
\draftthree{where the coefficients $c$ and $d$ are derived from linear fits to the relationship between $\text{log} M(H_2)$ and $Variable$, where the latter is chosen to be $M_\star$. The sample used to derive this linear fit consists of all HRS galaxies with \HI\ deficiencies $\leq$0.4, and upper limits are treated as measurements. An extensive description of the calibration of this H$_2$ deficiency parameter can be found in \S4.1 in \cite{Boselli2014}.}

A comparison between H$_2$ deficiencies from that work and those derived here is shown in Figure \ref{fig:def_verti_hera}. H$_2$ deficiencies of the VERTICO sample \draftthree{calculated using Equation \ref{eq:def_h2}} are systematically lower than those of the HRS sample by $\sim$0.25 dex, and \draftthree{by} $\sim$0.5 dex \draftthree{in case of} the two galaxies with the largest H$_2$ deficiences. \draftthree{There is a systematic offset in molecular gas masses in VERTICO and published molecular gas masses for the same objects in \citet{Boselli2014}, which is the result of differences in the calibration approach (see \S4.3 in B21). This likely causes differences in the derived H$_2$ deficiencies. Moreover, the control samples used to calibrate the relation are significantly different.} \citet{Boselli2014} calibrate the relation against a sub-sample of the HRS consisting of 101 spiral galaxies for which CO data \draftfour{are} available, and which have \HI\ deficiencies $\leq$ 0.4 dex. This \draftthree{method} could introduce a bias towards gas-rich galaxies, which would result in an overestimation of H$_2$ deficiencies for the Virgo systems. Since our def$_{\text{\HI,} M_\star}$ parameter agrees well with the def$_{\text{\HI,} R_\star}$ parameter from \citet[see also Figure \ref{fig:hi_def_vs_hi_def}]{Chung2009}, and H$_2$ deficiency was derived similarly, we are confident that the H$_2$ deficiencies \draftfour{we derive} are \draftthree{suitable for the purpose of this work}.

\startlongtable
\begin{deluxetable*}{l@{\hspace{1.8mm}}r@{\hspace{1.8mm}}r@{\hspace{0.1mm}}r@{\hspace{1.8mm}}r@{\hspace{1mm}}c@{\hspace{1mm}}c@{\hspace{1mm}}r@{\hspace{1.8mm}}r@{\hspace{1.8mm}}r@{\hspace{1.8mm}}r@{\hspace{1.8mm}}r@{\hspace{1.8mm}}r}
\tablecaption{Cold gas properties of the VERTICO sample.}
\tablehead{\colhead{Galaxy} & \colhead{RA (J2000)} & \colhead{Dec (J2000)} & \colhead{$v_{\rm opt}$} & \colhead{log $M_\star$} & $R_\star$ & \colhead{cl.} & \colhead{log $M_{\text{\HI}}$} & \colhead{def$_{\text{\HI,} R_\star}$} & \colhead{def$_{\text{\HI,} M_\star}$} & \colhead{log $M_{\text{mol}}$} & \colhead{def$_{\text{H}_2}$} \\ 
\colhead{-} & \colhead{-} & \colhead{-} & \colhead{$\mathrm{km\,s^{-1}}$} & \colhead{$\text{M}_\odot$} & kpc & \colhead{-} & \colhead{$\text{M}_\odot$}& \colhead{dex} & \colhead{dex} & \colhead{$\text{M}_\odot$} & \colhead{dex} \\ 
\colhead{(1)} & \colhead{(2)} & \colhead{(3)} & \colhead{(4)} & \colhead{(5)} & \colhead{(6)} & \colhead{(7)} & \colhead{(8)} & \colhead{(9)} & \colhead{(10)} & \colhead{(11)} & \colhead{(12)}}
\startdata
IC3392 & $12^\mathrm{h}28^\mathrm{m}43.27^\mathrm{s}$ & $14^\circ59{}^\prime57.48{}^{\prime\prime}$ &1678& 9.51 $\pm$ 0.1 &8.6& \textsc{iii} & 7.63 $\pm$ 0.35 & 1.15 $\pm$ 0.12 & 1.50$^{+ 0.34 } _{-0.34 }$& 8.73 $\pm$ 0.02 & -0.13 $^{+ 0.02 } _{-0.02 }$ \\
IC3418 & $12^\mathrm{h}29^\mathrm{m}43.92^\mathrm{s}$ & $11^\circ24{}^\prime16.88{}^{\prime\prime}$ &38&8.37& - & - & $\leq$6.9 & - & $\geq 2.47$    & $\leq$7.44 & $\geq-0.003$ \\      
NGC4064 & $12^\mathrm{h}04^\mathrm{m}11.26^\mathrm{s}$ & $18^\circ26{}^\prime39.12{}^{\prime\prime}$ &1000& 9.47 $\pm$ 0.1 &18& \textsc{iii} & 7.6 $\pm$ 0.26 & 1.79 $\pm$ 0.2 & 1.50$^{+ 0.26 } _{-0.26 }$& 8.41 $\pm$ 0.02 & -0.22 $^{+ 0.02 } _{-0.11 }$ \\
NGC4189 & $12^\mathrm{h}13^\mathrm{m}47.47^\mathrm{s}$ & $13^\circ25{}^\prime34.68{}^{\prime\prime}$ &1995& 9.75 $\pm$ 0.1 &9.6& \textsc{0} & 8.75 $\pm$ 0.06 & 0.25 $\pm$ 0.04 & 0.63$^{+ 0.09 } _{-0.06 }$& 8.69 $\pm$ 0.01 & -0.24 $^{+ 0.07 } _{-0.01 }$ \\
NGC4192 & $12^\mathrm{h}13^\mathrm{m}48.58^\mathrm{s}$ & $14^\circ53{}^\prime57.12{}^{\prime\prime}$ &-118& 10.78 $\pm$ 0.1 &57& \textsc{iv} & 9.63 $\pm$ 0.04 & 0.51 $\pm$ 0.2 & 0.23$^{+ 0.05 } _{-0.05 }$& 9.26 $\pm$ 0.01 & 0.37 $^{+ 0.2 } _{-0.12 }$ \\
NGC4216 & $12^\mathrm{h}15^\mathrm{m}54.19^\mathrm{s}$ & $13^\circ08{}^\prime54.96{}^{\prime\prime}$ &30& 10.91 $\pm$ 0.1 &38& \textsc{iv} & 9.25 $\pm$ 0.09 & 0.76 $\pm$ 0.2 & 0.70$^{+ 0.09 } _{-0.1 }$& 8.96 $\pm$ 0.01 & 0.56 $^{+ 0.06 } _{-0.19 }$ \\
NGC4222 & $12^\mathrm{h}16^\mathrm{m}22.56^\mathrm{s}$ & $13^\circ18{}^\prime25.20{}^{\prime\prime}$ &225& 9.45 $\pm$ 0.2 &15& \textsc{0} & 8.81 $\pm$ 0.1 & 0.32 $\pm$ 0.04 & 0.28$^{+ 0.1 } _{-0.18 }$& 8.06 $\pm$ 0.03 & 0.11 $^{+ 0.04 } _{-0.31 }$ \\
NGC4254\tablenotemark{a} & $12^\mathrm{h}18^\mathrm{m}49.68^\mathrm{s}$ & $14^\circ25{}^\prime05.52{}^{\prime\prime}$ &2453& 10.52 $\pm$ 0.1 &17& \textsc{i} & 9.65 $\pm$ 0.04 & -0.1 $\pm$ 0.02 & -0.03$^{+ 0.04 } _{-0.08 }$& 9.88 $\pm$ 0.00 & -0.51 $^{+ 0.12 } _{-0.02 }$ \\
NGC4293\tablenotemark{a} & $12^\mathrm{h}21^\mathrm{m}13.47^\mathrm{s}$ & $18^\circ23{}^\prime03.12{}^{\prime\prime}$ &717& 10.5 $\pm$ 0.1 &19& \textsc{iii} & 7.43 $\pm$ 0.48 & 2.25 $\pm$ 0.2 & 2.16$^{+ 0.48 } _{-0.49 }$& 8.86 $\pm$ 0.01 & 0.49 $^{+ 0.09 } _{-0.02 }$ \\
NGC4294 & $12^\mathrm{h}21^\mathrm{m}17.81^\mathrm{s}$ & $11^\circ30{}^\prime39.24{}^{\prime\prime}$ &421& 9.38 $\pm$ 0.1 &17& \textsc{i} & 9.21 $\pm$ 0.03 & -0.11 $\pm$ 0.02 & -0.18$^{+ 0.04 } _{-0.13 }$& 7.75 $\pm$ 0.07 & 0.28 $^{+ 0.1 } _{-0.2 }$ \\
NGC4298\tablenotemark{a} & $12^\mathrm{h}21^\mathrm{m}33.12^\mathrm{s}$ & $14^\circ36{}^\prime19.80{}^{\prime\prime}$ &1122& 10.11 $\pm$ 0.1 &15& \textsc{ii} & 8.69 $\pm$ 0.08 & 0.41 $\pm$ 0.02 & 0.68$^{+ 0.1 } _{-0.18 }$& 9.08 $\pm$ 0.01 & -0.12 $^{+ 0.06 } _{-0.05 }$ \\
NGC4299 & $12^\mathrm{h}21^\mathrm{m}40.71^\mathrm{s}$ & $11^\circ30{}^\prime06.12{}^{\prime\prime}$ &209& 9.06 $\pm$ 0.1 &8.6& \textsc{i} & 9.04 $\pm$ 0.02 & -0.43 $\pm$ 0.02 & -0.72$^{+ 0.53 } _{-0.3 }$& 7.58 $\pm$ 0.08 & -0.38 $^{+ 0.17 } _{-0.17 }$ \\
NGC4302 & $12^\mathrm{h}21^\mathrm{m}42.24^\mathrm{s}$ & $14^\circ35{}^\prime57.12{}^{\prime\prime}$ &1111& 10.47 $\pm$ 0.1 &26& \textsc{ii} & 9.17 $\pm$ 0.07 & 0.39 $\pm$ 0.02 & 0.41$^{+ 0.07 } _{-0.11 }$& 9.09 $\pm$ 0.01 & 0.24 $^{+ 0.05 } _{-0.02 }$ \\
NGC4321\tablenotemark{a} & $12^\mathrm{h}22^\mathrm{m}54.77^\mathrm{s}$ & $15^\circ49{}^\prime33.24{}^{\prime\prime}$ &1579& 10.71 $\pm$ 0.1 &31& \textsc{0} & 9.46 $\pm$ 0.02 & 0.35 $\pm$ 0.12 & 0.36$^{+ 0.04 } _{-0.02 }$& 9.84 $\pm$ 0.00 & -0.16 $^{+ 0.17 } _{-0.01 }$ \\
NGC4330 & $12^\mathrm{h}23^\mathrm{m}16.95^\mathrm{s}$ & $11^\circ22{}^\prime04.08{}^{\prime\prime}$ &1567& 9.63 $\pm$ 0.15 &18& \textsc{ii} & 8.65 $\pm$ 0.1 & 0.8 $\pm$ 0.04 & 0.64$^{+ 0.11 } _{-0.1 }$& 8.34 $\pm$ 0.03 & -0.01 $^{+ 0.03 } _{-0.04 }$ \\
NGC4351 & $12^\mathrm{h}24^\mathrm{m}01.30^\mathrm{s}$ & $12^\circ12{}^\prime15.12{}^{\prime\prime}$ &2388& 9.37 $\pm$ 0.1 &5.1& \textsc{i} & 8.48 $\pm$ 0.06 & 0.23 $\pm$ 0.02 & 0.55$^{+ 0.06 } _{-0.14 }$& 7.85 $\pm$ 0.04 & 0.14 $^{+ 0.11 } _{-0.18 }$ \\
NGC4380 & $12^\mathrm{h}25^\mathrm{m}22.16^\mathrm{s}$ & $10^\circ01{}^\prime00.12{}^{\prime\prime}$ &935& 10.11 $\pm$ 0.1 &11& \textsc{iv} & 8.1 $\pm$ 0.19 & 1.13 $\pm$ 0.2 & 1.27$^{+ 0.2 } _{-0.25 }$& 8.59 $\pm$ 0.01 & 0.37 $^{+ 0.06 } _{-0.05 }$ \\
NGC4383 & $12^\mathrm{h}25^\mathrm{m}25.47^\mathrm{s}$ & $16^\circ28{}^\prime11.64{}^{\prime\prime}$ &1663& 9.44 $\pm$ 0.1 &6.4& \textsc{0} & 9.47 $\pm$ 0.05 & -0.81 $\pm$ 0.2 & -0.39$^{+ 0.05 } _{-0.07 }$& 8.37 $\pm$ 0.02 & -0.21 $^{+ 0.02 } _{-0.17 }$ \\
NGC4388 & $12^\mathrm{h}25^\mathrm{m}46.61^\mathrm{s}$ & $12^\circ39{}^\prime46.44{}^{\prime\prime}$ &2538& 10.07 $\pm$ 0.1 &26& \textsc{ii} & 8.57 $\pm$ 0.26 & 1.16 $\pm$ 0.12 & 0.84$^{+ 0.28 } _{-0.3 }$& 8.89 $\pm$ 0.01 & 0 $^{+ 0.1 } _{-0.03 }$ \\
NGC4394 & $12^\mathrm{h}25^\mathrm{m}55.66^\mathrm{s}$ & $18^\circ12{}^\prime52.20{}^{\prime\prime}$ &772& 10.34 $\pm$ 0.1 &18& \textsc{iv} & 8.64 $\pm$ 0.03 & 0.62 $\pm$ 0.2 & 0.90$^{+ 0.07 } _{-0.03 }$& 7.98 $\pm$ 0.04 & 1.26 $^{+ 0.05 } _{-0.04 }$ \\
NGC4396 & $12^\mathrm{h}25^\mathrm{m}59.66^\mathrm{s}$ & $15^\circ40{}^\prime10.20{}^{\prime\prime}$ &-115& 9.36 $\pm$ 0.1 &19& \textsc{i} & 8.94 $\pm$ 0.08 & 0.3 $\pm$ 0.04 & 0.08$^{+ 0.09 } _{-0.15 }$& 8.05 $\pm$ 0.04 & -0.1 $^{+ 0.14 } _{-0.17 }$ \\
NGC4402\tablenotemark{a} & $12^\mathrm{h}26^\mathrm{m}07.34^\mathrm{s}$ & $13^\circ06{}^\prime45.00{}^{\prime\prime}$ &190& 10.13 $\pm$ 0.1 &26& \textsc{ii} & 8.57 $\pm$ 0.18 & 0.74 $\pm$ 0.12 & 0.82$^{+ 0.19 } _{-0.23 }$& 9.10 $\pm$ 0.01 & -0.1 $^{+ 0.04 } _{-0.07 }$ \\
NGC4405 & $12^\mathrm{h}26^\mathrm{m}07.11^\mathrm{s}$ & $16^\circ10{}^\prime51.60{}^{\prime\prime}$ &1751& 9.75 $\pm$ 0.1 &8.3& \textsc{iii} & 7.65 $\pm$ 0.26 & 0.95 $\pm$ 0.2 & 1.73$^{+ 0.27 } _{-0.26 }$& 8.27 $\pm$ 0.02 & 0.18 $^{+ 0.07 } _{-0.02 }$ \\
NGC4419 & $12^\mathrm{h}26^\mathrm{m}56.35^\mathrm{s}$ & $15^\circ02{}^\prime51.36{}^{\prime\prime}$ &-228& 10.06 $\pm$ 0.1 &13& \textsc{iii} & 7.76 $\pm$ 0.63 & 1.37 $\pm$ 0.2 & 1.65$^{+ 0.63 } _{-0.63 }$& 9.10 $\pm$ 0.00 & -0.23 $^{+ 0.09 } _{-0.02 }$ \\
NGC4424\tablenotemark{a} & $12^\mathrm{h}27^\mathrm{m}11.69^\mathrm{s}$ & $09^\circ25{}^\prime14.16{}^{\prime\prime}$ &447& 9.89 $\pm$ 0.1 &11& \textsc{ii} & 8.28 $\pm$ 0.07 & 0.97 $\pm$ 0.2 & 1.14$^{+ 0.1 } _{-0.1 }$& 8.31 $\pm$ 0.01 & 0.35 $^{+ 0.03 } _{-0.06 }$ \\
NGC4450 & $12^\mathrm{h}28^\mathrm{m}29.23^\mathrm{s}$ & $17^\circ05{}^\prime04.56{}^{\prime\prime}$ &2048& 10.7 $\pm$ 0.1 &22& \textsc{iv} & 8.45 $\pm$ 0.08 & 1.17 $\pm$ 0.2 & 1.35$^{+ 0.09 } _{-0.08 }$& 8.67 $\pm$ 0.01 & 1 $^{+ 0.16 } _{-0.03 }$ \\
NGC4457\tablenotemark{a} & $12^\mathrm{h}28^\mathrm{m}59.02^\mathrm{s}$ & $03^\circ34{}^\prime14.16{}^{\prime\prime}$ &738& 10.42 $\pm$ 0.1 &12& \textsc{iii} & 8.29 $\pm$ 0.11 & 0.92 $\pm$ 0.2 & 1.29$^{+ 0.13 } _{-0.12 }$& 9.02 $\pm$ 0.01 & 0.27 $^{+ 0.02 } _{-0.03 }$ \\
NGC4501 & $12^\mathrm{h}31^\mathrm{m}59.33^\mathrm{s}$ & $14^\circ25{}^\prime10.92{}^{\prime\prime}$ &2120& 11 $\pm$ 0.1 &35& \textsc{ii} & 9.22 $\pm$ 0.06 & 0.58 $\pm$ 0.12 & 0.80$^{+ 0.07 } _{-0.07 }$& 9.69 $\pm$ 0.00 & -0.03 $^{+ 0.15 } _{-0.05 }$ \\
NGC4522 & $12^\mathrm{h}33^\mathrm{m}39.72^\mathrm{s}$ & $09^\circ10{}^\prime26.76{}^{\prime\prime}$ &2332& 9.66 $\pm$ 0.1 &16& \textsc{ii} & 8.53 $\pm$ 0.13 & 0.86 $\pm$ 0.02 & 0.79$^{+ 0.13 } _{-0.13 }$& 8.30 $\pm$ 0.02 & 0.06 $^{+ 0.02 } _{-0.04 }$ \\
NGC4532 & $12^\mathrm{h}34^\mathrm{m}19.35^\mathrm{s}$ & $06^\circ28{}^\prime05.52{}^{\prime\prime}$ &2154& 9.25 $\pm$ 0.1 &8& \textsc{0} & 9.29 $\pm$ 0.03 & -0.06 $\pm$ 0.06 & -0.25$^{+ 0.12 } _{-0.06 }$& 8.25 $\pm$ 0.02 & -0.58 $^{+ 0.15 } _{-0.15 }$ \\
NGC4533 & $12^\mathrm{h}34^\mathrm{m}22.03^\mathrm{s}$ & $02^\circ19{}^\prime33.24{}^{\prime\prime}$ &1753& 9.2 $\pm$ 0.1 &9.6& - & 8.43 $\pm$ 0.11 & 0.51 $\pm$ 0.04 & 0.59$^{+ 0.13 } _{-0.37 }$& 6.94 $\pm$ 0.26 & 0.6 $^{+ 0.3 } _{-0.3 }$ \\
NGC4535\tablenotemark{a} & $12^\mathrm{h}34^\mathrm{m}20.26^\mathrm{s}$ & $08^\circ11{}^\prime53.52{}^{\prime\prime}$ &1973& 10.49 $\pm$ 0.1 &29& \textsc{i} & 9.52 $\pm$ 0.02 & 0.41 $\pm$ 0.12 & 0.07$^{+ 0.02 } _{-0.1 }$& 9.41 $\pm$ 0.01 & -0.06 $^{+ 0.08 } _{-0.02 }$ \\
NGC4536\tablenotemark{a} & $12^\mathrm{h}34^\mathrm{m}27.12^\mathrm{s}$ & $02^\circ11{}^\prime16.08{}^{\prime\prime}$ &1894& 10.19 $\pm$ 0.1 &42& \textsc{0} & 9.68 $\pm$ 0.02 & 0.16 $\pm$ 0.12 & -0.26$^{+ 0.03 } _{-0.08 }$& 9.35 $\pm$ 0.01 & -0.24 $^{+ 0.01 } _{-0.09 }$ \\
NGC4548\tablenotemark{a} & $12^\mathrm{h}35^\mathrm{m}26.64^\mathrm{s}$ & $14^\circ29{}^\prime43.80{}^{\prime\prime}$ &498& 10.65 $\pm$ 0.1 &26& \textsc{iv} & 8.81 $\pm$ 0.03 & 0.82 $\pm$ 0.12 & 0.96$^{+ 0.04 } _{-0.03 }$& 8.99 $\pm$ 0.01 & 0.63 $^{+ 0.07 } _{-0.09 }$ \\
NGC4561 & $12^\mathrm{h}36^\mathrm{m}08.14^\mathrm{s}$ & $19^\circ19{}^\prime21.72{}^{\prime\prime}$ &1441& 9.09 $\pm$ 0.1 &6.1& \textsc{0} & 9.15 $\pm$ 0.03 & -0.71 $\pm$ 0.02 & -0.64$^{+ 0.41 } _{-0.46 }$& 7.31 $\pm$ 0.19 & -0.04 $^{+ 0.24 } _{-0.24 }$ \\
NGC4567 & $12^\mathrm{h}36^\mathrm{m}33.07^\mathrm{s}$ & $11^\circ15{}^\prime29.16{}^{\prime\prime}$ &2213& 10.25 $\pm$ 0.1 &12& \textsc{0} & 8.97 $\pm$ 0.03 & 0.13 $\pm$ 0.12 & -0.27$^{+ 0.03 } _{-0.05 }$& 8.84 $\pm$ 0.00 & 0.33 $^{+ 0.02 } _{-0.02 }$ \\
NGC4568 & $12^\mathrm{h}36^\mathrm{m}34.34^\mathrm{s}$ & $11^\circ14{}^\prime21.84{}^{\prime\prime}$ &2260& 10.47 $\pm$ 0.1 &22& \textsc{0} & 9.18 $\pm$ 0.05 & 0.38 $\pm$ 0.12 & 0.40$^{+ 0.05 } _{-0.1 }$& 9.41 $\pm$ 0.00 & -0.08 $^{+ 0.05 } _{-0.02 }$ \\
NGC4569\tablenotemark{a} & $12^\mathrm{h}36^\mathrm{m}50.12^\mathrm{s}$ & $13^\circ09{}^\prime55.08{}^{\prime\prime}$ &-220& 10.86 $\pm$ 0.1 &50& \textsc{iii} & 8.79 $\pm$ 0.1 & 1.47 $\pm$ 0.2 & 1.12$^{+ 0.11 } _{-0.11 }$& 9.58 $\pm$ 0.00 & -0.23 $^{+ 0.08 } _{-0.19 }$ \\
NGC4579\tablenotemark{a} & $12^\mathrm{h}37^\mathrm{m}43.44^\mathrm{s}$ & $11^\circ49{}^\prime05.52{}^{\prime\prime}$ &1627& 10.92 $\pm$ 0.1 &21& \textsc{iv} & 8.75 $\pm$ 0.12 & 0.95 $\pm$ 0.2 & 1.20$^{+ 0.12 } _{-0.12 }$& 9.31 $\pm$ 0.00 & 0.22 $^{+ 0.08 } _{-0.17 }$ \\
NGC4580 & $12^\mathrm{h}37^\mathrm{m}48.38^\mathrm{s}$ & $05^\circ22{}^\prime06.24{}^{\prime\prime}$ &1227& 9.94 $\pm$ 0.1 &8.6& \textsc{iii} & 7.45 $\pm$ 0.34 & 1.53 $\pm$ 0.2 & 1.99$^{+ 0.36 } _{-0.35 }$& 8.55 $\pm$ 0.01 & 0.17 $^{+ 0.03 } _{-0.03 }$ \\
NGC4606 & $12^\mathrm{h}40^\mathrm{m}57.62^\mathrm{s}$ & $11^\circ54{}^\prime43.56{}^{\prime\prime}$ &1653& 9.61 $\pm$ 0.1 &13& \textsc{iii} & 7.4 $\pm$ 0.22 & 1.64 $\pm$ 0.2 & 1.86$^{+ 0.23 } _{-0.23 }$& 8.14 $\pm$ 0.02 & 0.18 $^{+ 0.02 } _{-0.02 }$ \\
NGC4607 & $12^\mathrm{h}41^\mathrm{m}12.39^\mathrm{s}$ & $11^\circ53{}^\prime09.60{}^{\prime\prime}$ &2284& 9.64 $\pm$ 0.1 &12& \textsc{iii} & 8.34 $\pm$ 0.16 & 0.82 $\pm$ 0.12 & 0.96$^{+ 0.16 } _{-0.16 }$& 8.58 $\pm$ 0.01 & -0.24 $^{+ 0.01 } _{-0.03 }$ \\
NGC4651 & $12^\mathrm{h}43^\mathrm{m}42.72^\mathrm{s}$ & $16^\circ23{}^\prime37.68{}^{\prime\prime}$ &788& 10.31 $\pm$ 0.1 &13& \textsc{0} & 9.61 $\pm$ 0.03 & -0.3 $\pm$ 0.02 & 0.09$^{+ 0.04 } _{-0.03 }$& 8.67 $\pm$ 0.01 & -0.46 $^{+ 0.04 } _{-0.01 }$ \\
NGC4654\tablenotemark{a} & $12^\mathrm{h}43^\mathrm{m}56.76^\mathrm{s}$ & $13^\circ07{}^\prime32.52{}^{\prime\prime}$ &1035& 10.26 $\pm$ 0.1 &12& \textsc{ii} & 9.47 $\pm$ 0.03 & 0.12 $\pm$ 0.02 & -0.01$^{+ 0.03 } _{-0.04 }$& 9.33 $\pm$ 0.00 & -0.15 $^{+ 0.03 } _{-0.01 }$ \\
NGC4689\tablenotemark{a} & $12^\mathrm{h}47^\mathrm{m}45.68^\mathrm{s}$ & $13^\circ45{}^\prime42.12{}^{\prime\prime}$ &1522& 10.16 $\pm$ 0.1 &19& \textsc{iv} & 8.67 $\pm$ 0.05 & 0.68 $\pm$ 0.12 & 0.73$^{+ 0.06 } _{-0.12 }$& 9.02 $\pm$ 0.01 & 0.04 $^{+ 0.01 } _{-0.09 }$ \\
NGC4694\tablenotemark{a} & $12^\mathrm{h}48^\mathrm{m}15.08^\mathrm{s}$ & $10^\circ59{}^\prime00.60{}^{\prime\prime}$ &1211& 9.94 $\pm$ 0.1 &19& \textsc{ii} & 8.4 $\pm$ 0.03 & 0.83 $\pm$ 0.2 & 1.03$^{+ 0.12 } _{-0.08 }$& 8.23 $\pm$ 0.02 & 0.5 $^{+ 0.03 } _{-0.04 }$ \\
NGC4698 & $12^\mathrm{h}48^\mathrm{m}22.99^\mathrm{s}$ & $08^\circ29{}^\prime15.00{}^{\prime\prime}$ &1032& 10.49 $\pm$ 0.1 &19& \textsc{i} & 9.21 $\pm$ 0.03 & 0.02 $\pm$ 0.2 & 0.37$^{+ 0.04 } _{-0.1 }$& 7.94 $\pm$ 0.10 & 1.4 $^{+ 0.12 } _{-0.1 }$ \\
NGC4713 & $12^\mathrm{h}49^\mathrm{m}57.65^\mathrm{s}$ & $05^\circ18{}^\prime39.60{}^{\prime\prime}$ &631& 9.31 $\pm$ 0.16 &10& \textsc{0} & 9.46 $\pm$ 0.03 & -0.31 $\pm$ 0.04 & -0.42$^{+ 0.1 } _{-0.03 }$& 8.38 $\pm$ 0.02 & -0.56 $^{+ 0.22 } _{-0.24 }$ \\
NGC4772 & $12^\mathrm{h}53^\mathrm{m}29.12^\mathrm{s}$ & $02^\circ10{}^\prime06.24{}^{\prime\prime}$ &1042& 10.18 $\pm$ 0.1 &9.9& \textsc{0} & 8.92 $\pm$ 0.06 & 0.15 $\pm$ 0.2 & 0.48$^{+ 0.06 } _{-0.11 }$& 7.40 $\pm$ 0.27 & 1.7 $^{+ 0.27 } _{-0.29 }$ \\
NGC4808 & $12^\mathrm{h}55^\mathrm{m}48.94^\mathrm{s}$ & $04^\circ18{}^\prime15.12{}^{\prime\prime}$ &738& 9.63 $\pm$ 0.1 &11& \textsc{0} & 9.55 $\pm$ 0.03 & -0.58 $\pm$ 0.04 & -0.27$^{+ 0.03 } _{-0.04 }$& 8.74 $\pm$ 0.01 & -0.41 $^{+ 0.01 } _{-0.03 }$ \\
VCC1581 & $12^\mathrm{h}34^\mathrm{m}45.30^\mathrm{s}$ & $06^\circ18{}^\prime02.00{}^{\prime\prime}$ &2039& 8.47 $\pm$ 0.1 & - & - & 8.5 $\pm$ 0.04 & -0.06 $\pm$ 0.06 & 0.62$^{+ 0.36 } _{-0.36 }$& $\leq$6.84 & $\geq$ 0.69 \\     
\enddata
\tablecomments{\draftone{Columns are: (1) common name; 
(2) right ascension (J2000) of the galaxy optical centre; (3) declination (J2000) of the galaxy optical centre; 
(4) optical heliocentric recession velocity; (5) stellar mass, (see \S \ref{subsub:stellar_masses}); \drafttwo{(6) Optical radius, defined as the isodensity radius of the stellar mass \draftthree{distribution}, where $\Sigma_\star = 1\ \text{M}_\odot\ \text{pc}^{-2}$ (\S \ref{subsub:stellar_masses}). The uncertainty is dominated by the resolution of the stellar mass surface density maps, and therefore equals 0.3 kpc for all galaxies}; (7) \HI\ classification from VIVA (see \S \ref{sub:hi_classes}); (8) \HI\ mass, from table 3 in \citet{Chung2009}; (9) \HI\ deficiency, from table 3 in \citet[see also \S \ref{sub:HI_deficiency}]{Chung2009}; (10) \HI\ deficiency, calculated using the predicted \HI\ mass from field galaxies at fixed stellar mass (see \S \ref{sub:HI_deficiency}); (11) molecular gas mass, from table 2 in B21; (12) molecular gas deficiency, see \S \ref{sub:H2_deficiency}. Columns 2--4 are drawn from the NASA/IPAC Extragalactic Database \refrep{\citep{NED}}.}
}
\tablenotetext{a}{The ALMA observations of these galaxies are from the archive; see \S \ref{sec:sample} for more details.}
\label{tab:VERTICO-sample}
\end{deluxetable*}

\subsubsection{Stellar masses \& radii} \label{subsub:stellar_masses}
Stellar masses for the VERTICO sample are adopted from the $z=0$ Multi-wavelength Galaxy Synthesis ($z$0MGS, \citealt{Leroy2019}), \draftthree{who use the initial mass function (IMF) from \citet{Kroupa2003}}. The only exception is IC3418, which is not included in $z$0MGS. For this galaxy we adopt a stellar mass of $M_\star = 10^{8.37}\ \text{M}_\odot$, following \citet[see also \S 2.2 in B21]{Fumagalli2011}. \draftone{Stellar radii are here defined as the isodensity radius with a threshold of $\Sigma_\star$ = 1 M$_\odot\ \text{pc}^{-2}$. \drafttwo{These radii are measured by identifying the outermost annulus or slice of the stellar mass radial profile (measured as described in \S \ref{sub:rad_profs}) that is still \draftthree{above} this threshold, and adopting its radius. \draftfour{For consistency with the gas radius and surface density measurements (see \S \ref{sub:rad_profs}) the stellar surface density maps are not corrected for inclination before radial profiles are derived.} The uncertainty in the radius is a combination of the resolution of the stellar mass surface density maps (9$^{\prime\prime}$, corresponding to $\sim$720 pc at the distance of the Virgo cluster) and the uncertainty in the stellar mass surface density.}}

\drafttwo{We produce stellar mass surface density maps from Wide-field Infrared Survey Explorer (WISE) band 1 photometry, following the procedure laid out in \citet[\draftthree{and therefore using the same IMF as that used for the stellar masses, see above}]{Leroy2019}. All images are convolved from their native resolution to a 9$^{\prime\prime}$ Gaussian beam, using the convolution kernels from \cite{Aniano2011}. All Gaia DR2 stars within the image area are masked. Image backgrounds are estimated and subtracted with the \texttt{Background2D} function from \texttt{Astropy}. For each pixel we determine the local mass-to-light ratio (at 3.4 $\mu$m) using the WISE band 3 to WISE band 1 colour as an `sSFR-like' proxy, and following the calibrations given in the Appendix of \citet{Leroy2019}. The WISE band 1 images are then combined with the derived mass-to-light ratios to produce resolved
stellar mass surface density maps in units of M$_\odot$ pc$^{-2}$.}

\subsection{Radial profiles} \label{sub:rad_profs}
H$_2$ radial profiles are calculated as described in \S 4.3 of B21. In summary, they reflect the azimuthally averaged integrated H$_2$ surface densities in elliptical annuli overlaid on the moment 0 maps \drafttwo{at native resolution} (these maps can be seen in Appendix A of B21). \draftone{While some authors opt to correct for inclination in order to obtain a measure of the intrinsic surface density (assuming the gas is distributed in a flat disk), here we do not apply this correction, due to the disturbed nature of several of the sources.} \refrep{Similarly, it should be kept in mind that any extraplanar gas that may be present in these disturbed sources could be assigned to artificially large radii due to projection effects, especially in highly inclined galaxies. This would result in a flattened radial profile, and possibly an overestimation of the radius of the molecular gas ``disc''.} 

Pixels that do not contain any \drafttwo{detected} emission are included \drafttwo{as zeros}. This means that the surface density at each radius represents the average surface density in the \emph{entire} corresponding annulus, rather than the average surface density of the molecular gas detected inside the annulus. Therefore, \draftfour{on} the outskirts of the molecular gas discs, where CO is not detected in all pixels, this can result in low average molecular gas surface densities. This effect will be stronger in galaxies with very asymmetric molecular gas discs. \drafttwo{We continue to add annuli until there are no more detected pixels \draftthree{in} the outer annulus. Since we are working with clipped moment 0 maps, all non-zero pixels in the map are significant.}

\drafttwo{To allow for accurate interpolation in our calculation of the radial profiles, we use annuli with widths of one pixel \drafttwo{(2$^{\prime\prime}$, equivalent to $\sim$ 160 pc at the distance of Virgo)} across the minor axis. This approach differs from that of B21, who use annuli with widths of one beam \drafttwo{(7 \draftthree{--} 10$^{\prime\prime}$; B21, table 2)} across the minor axis. To allow for a fair comparison between galaxies, we then interpolate the resulting radii \drafttwo{($R$)} to match certain fractions of the galaxy's stellar radius (at each $\Delta 0.2 R_\star$).} This means that in some cases $\Delta R$ can be smaller than one beam, in which case the integrated H$_2$ surface densities in the corresponding annuli are not independent. Since we are interested in the relative shapes of the profiles, and their corresponding radii and \drafttwo{average surface densities}, this \drafttwo{will not impact our analysis.}

\draftfour{The eccentricities of the annuli, along with the inclinations and position angles of the galaxies, are listed in table 1 of B21.} For highly\drafttwo{-}inclined ($i\gtrsim80^o$) galaxies we take slices along the major axis instead \draftthree{of annuli} (extending infinitely in the direction of the minor axis). \draftthree{The emission in the corresponding slices on each side of the galactic centre (at the same galactocentric radius) is then averaged to obtain the radial profile (see also B21).} 

Molecular gas radial profiles of all \drafttwo{CO-}detected VERTICO galaxies are shown in B21; figure 9, and in \draftthree{Figure \ref{fig:rad_profs}} along with their \HI\ radial profiles \draftone{(which were calculated in the same way, \draftfour{i.e. by placing annuli on the observed (inclination-uncorrected) \HI\ surface density maps})}, as well as \draftthree{the ratio between both profiles}. \drafttwo{For this Figure the 15$^{\prime\prime}$ resolution VERTICO moment 0 maps were used, to match the \HI\ data.}

\subsection{H$_2$ radii} \label{sub:radii}
\draftthree{In this paper}, the radius of the H$_2$ disc is defined as its isodensity radius: the radius at which the \draftfour{observed} surface density of the molecular gas disc drops below a certain threshold value (henceforth referred to as the ``isodensity threshold''). In cases where the surface density is not described by a strictly declining function (for example when a galaxy has pronounced spiral arms and inter-arm regions with low molecular gas surface densities), it is possible that this \draftthree{threshold is reached} at multiple radii. In such cases, the outermost radius is defined to be the isodensity radius. 

Typically, the radius of the H$_2$ disc is defined at isodensity thresholds of $\sim 5\  \text{M}_\odot\ \text{pc}^{-2}$, where the molecular gas starts dominating the cold gas reservoir \drafttwo{(e.g. \citealt{Walter2008, Leroy2008})}. However, depending on which part of the disc we are interested in (e.g. the core or the very outskirts), it is sometimes useful to use lower or higher isodensity thresholds. Throughout this work, \draftthree{we specify} which isodensity threshold is used to define the radius of the molecular gas disc \draftthree{for a particular investigation} \draftone{(e.g. in Figure \ref{fig:H2_radii})}.

Because the main goal of this work is to compare different galaxies, and because we are interested in the \emph{relative} size of the molecular gas disc, H$_2$ radii are normalised by the radius of the stellar disc \draftone{(see \S \ref{subsub:stellar_masses})}.

\subsection{Median H$_2$ surface densities} \label{sub:densities}
The \draftone{median} H$_2$ surface density is here defined as the median H$_2$ surface density of the moment 0 map within \draftthree{an elliptical aperture of which the semi-major axis is a certain \draftfour{observed} fraction of $R_\star$.} For highly\drafttwo{-}inclined galaxies and galaxies with very little, flocculent CO emission, \draftone{such a median} \draftthree{surface} density is not well-defined. For this reason, we do not \draftthree{measure it} for the following galaxies: NGC4192, NGC4216, NGC4222, NGC4299, NGC4302, NGC4330, NGC4388, NGC4396, NGC4402, NGC4533, NGC4607, NGC4698, and NGC4772. \draftthree{As a result,} \drafttwo{median} H$_2$ surface densities are \draftthree{measured} for 34/49 detected VERTICO galaxies. \draftone{Median H$_2$ surface densities will be studied as a function of \HI\ and H$_2$ deficiency (see \S \ref{sub:h2_surf_dens} and Figure \ref{fig:H2_densities}).}

\subsection{Classification of \HI\ removal stages} \label{sub:hi_classes}
\begin{figure*}
	\centering
	\subfloat
	{\includegraphics[height=0.442\textwidth]{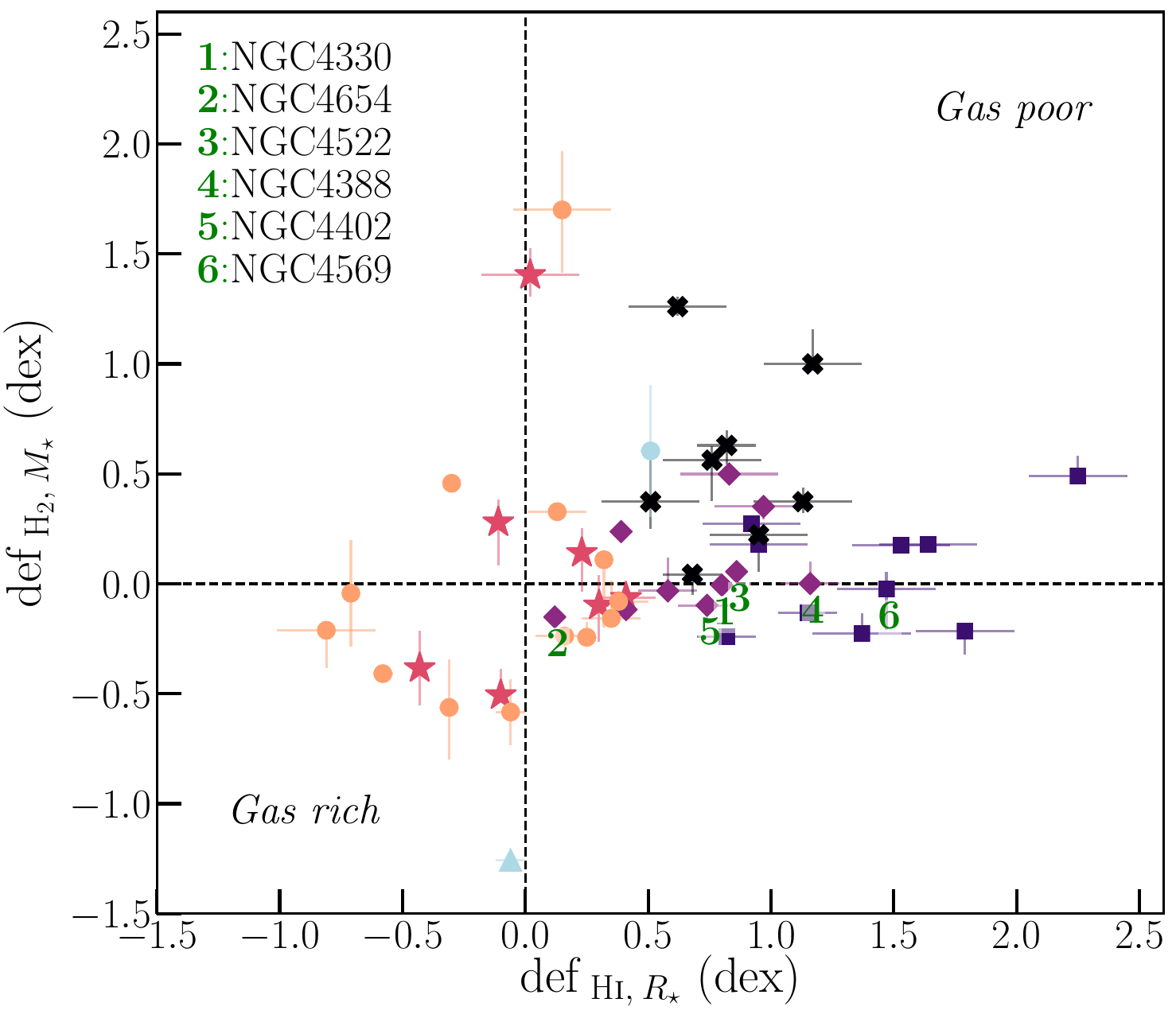}} \hspace{0.5mm}
	\subfloat
	{\includegraphics[width=0.475\textwidth, height=0.442\textwidth]{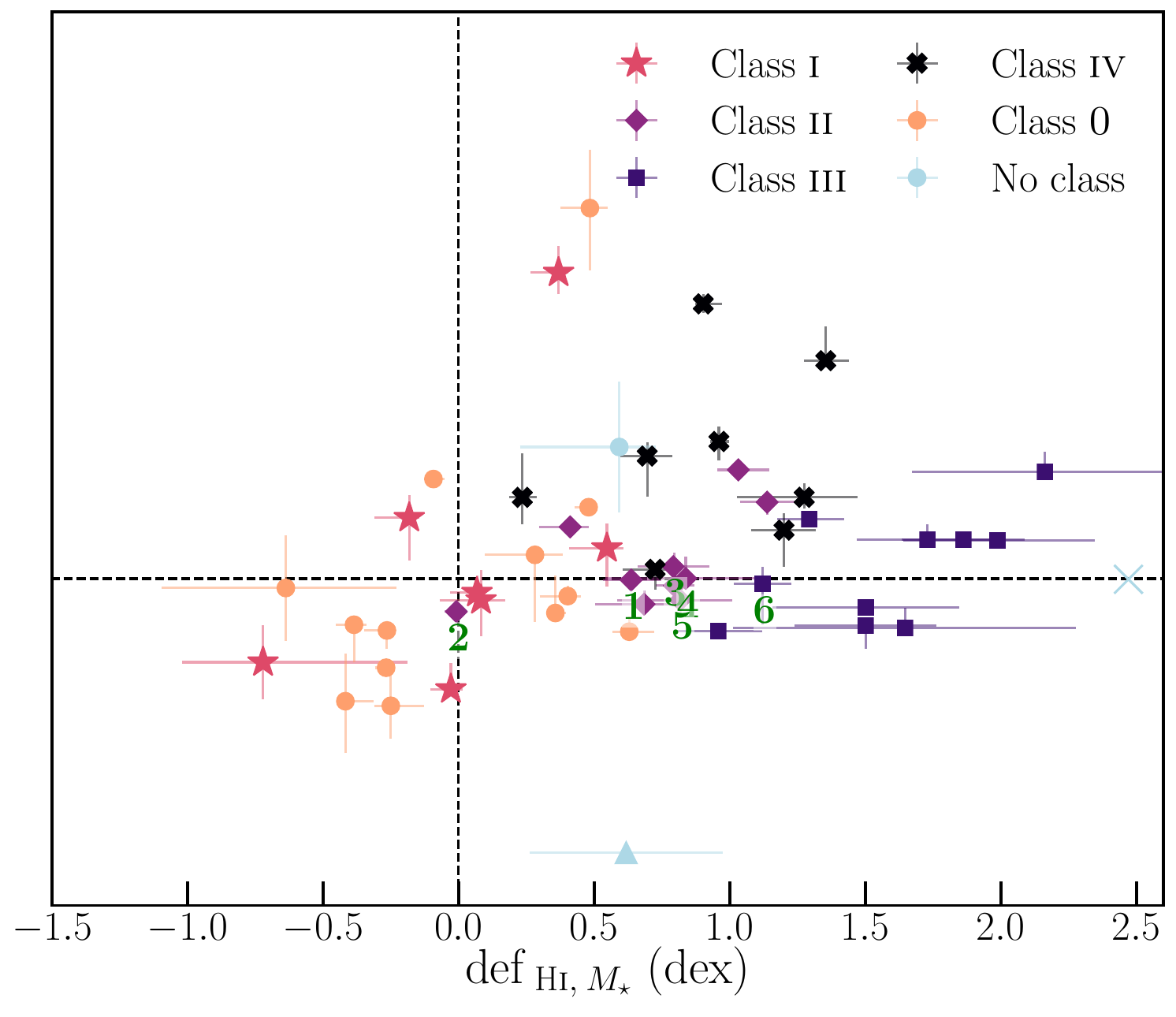}}
	\caption{\draftone{H$_2$ deficiency as a function of \HI\ deficiency, where \HI\ deficiency is estimated in two different ways (def$_{\text{\HI, }R_\star}$, left-hand panel, and def$_{\text{\HI, }M_\star}$, right-hand panel). Classes refer to the \HI\ classification from \citet[\draftthree{see \S \ref{sub:hi_classes}}]{Yoon2017}, based on deficiency and morphology. \draftthree{VCC1581, which was not detected in CO, is represented by the light-blue triangle, indicating the lower limit on its molecular gas deficiency.} The x-shaped marker in the right panel indicates \refrep{IC3418}, which is a lower limit both in \HI\ and H$_2$ deficiency (it is not present in the left-hand panel because no limit on the \HI\ deficiency was provided by \citealt{Chung2009}). The numbered galaxies \draftfour{(green numbers are placed below the relevant markers)} represent well-studied ``smoking-gun'' RPS galaxies, listed in the top-left corner in order of pre-to-post RP peak. \draftfour{There is a weak correlation between \HI\ and H$_2$ deficiency:} the Spearman correlation coefficient is 0.28 $\pm$ 0.05, with a typical \textit{p}-value of 0.05, for the left-hand panel and 0.36 $\pm$ 0.04, with a typical \textit{p}-value of 0.01, for the right-hand panel. Most galaxies lie in the upper-right quadrant, and are thus both \HI\ and H$_2$ deficient. On average, galaxies that have \HI\ reservoirs similar to those of field galaxies (class \textsc{0} galaxies) are least H$_2$ deficient, while galaxies that are \HI-deficient because of their lower \HI\ surface densities (class \textsc{iv} galaxies) are most H$_2$ deficient. VERTICO galaxies undergoing RPS \draftthree{are relatively H$_2$-rich compared to other VERTICO galaxies with similar \HI\ deficiencies.}}}
	\label{fig:HI_H2_def}
\end{figure*}
\citet{Yoon2017} \drafttwo{study} in detail the \HI\ \refrep{properties} of the galaxies in VIVA, the sample on which the VERTICO sample is based. \refrep{They are particularly interested in \HI\ deficiency, radius, and morphology, since any gas stripping is most notably reflected in these parameters. Therefore, to study the effects of gas stripping as systematically as possible, they define a classification for the Virgo galaxies based on the combination these \HI\ properties.} In particular, each class spans a limited range of \HI\ deficiency and relative \draftthree{\HI} extent. This classification is described in detail in \S 2.1 of \citet{Yoon2017}. In summary, galaxies comprising each class are:
\begin{enumerate}
\item [Class \textsc{i}] Galaxies with a one-sided \HI\ feature, no truncation of the disc and a relatively symmetric stellar disc. \drafttwo{Their} \HI\ deficiency varies but is overall close to that in field galaxies.
\item [Class \textsc{ii}] Galaxies with highly asymmetric \HI\ discs, caused by tails and/or asymmetric truncation, deficient in \HI\ (average deficiency (def$_{\text{\HI,}R_\star}$) of $\sim$0.8 dex).
\item [Class \textsc{iii}] Galaxies with symmetric but severely truncated \HI\ discs that are extremely \HI\ deficient (average deficiency of $\gtrsim$1.4 dex).
\item [Class \textsc{iv}] Galaxies with symmetric \HI\ discs that are only marginally truncated, but with lower \HI\ surface density than the other classes, and an average deficiency of $\sim$0.8 dex (similar to class \textsc{ii}).
\item [Class \textsc{0}] Galaxies that do not fit in the above classes, including galaxies that resemble ``normal'' field spirals, extremely \HI-rich galaxies with extended \HI\ discs, galaxies that have asymmetric \HI\ discs but no truncation, showing clear signs of tidal interaction.
\end{enumerate}
\refrep{To build upon this work by \citet{Yoon2017}, and investigate if there is any correlation between their \HI\ classification and molecular gas deficiency, these classes are highlighted in Figure \ref{fig:HI_H2_def}.} Note that \drafttwo{the dwarf galaxy} NGC4533 is the only galaxy in the VERTICO sample for which no \HI\ class is \drafttwo{given in \citet{Yoon2017}}, \draftthree{who exclude 5 dwarf galaxies from the classification scheme.} The reason for this is that such dwarf galaxies \draftone{might be} more sensitive to their local environment rather than their orbital history, which was the main focus of the VIVA classification.

\section{Results} \label{sec:results}
\subsection{Do \HI\ deficiencies predict H$_2$ deficiencies?}
Figure \ref{fig:HI_H2_def} shows the relationship between \HI\ deficiency and H$_2$ deficiency. The relationship is shown for two different definitions of \HI\ deficiency: def$_{\text{\HI, }R_\star}$ in the left-hand panel, and def$_{\text{\HI, }M_\star}$ in the right-hand panel \draftthree{(see \S \ref{sub:HI_deficiency} for their respective definitions}). \drafttwo{\HI\ deficiencies from both definitions correlate well \draftfour{(the typical scatter is $\sim$0.2 dex, see Figure \ref{fig:hi_def_vs_hi_def}).}} The classes in the legend refer to the \HI\ classification described in \S \ref{sub:hi_classes}. Six well-studied cases of galaxies undergoing RPS are highlighted with red numbers 1-6, and listed in the top-left corner \draftone{of the left-hand panel} in order of pre-to-post RP peak \draftthree{\citep{Vollmer2003, Vollmer2009, Vollmer2004, Vollmer2008, Vollmer2012b, Boselli2006b, Boselli2016, Lee2017, Lee2018}}.

\draftthree{There is a weak correlation between \HI\ and H$_2$ deficiency. The Spearman correlation coefficient is 0.28 $\pm$ 0.05 in the left-hand panel (where def$_{\text{\HI}} \equiv$  def$_{\text{\HI, }R_\star}$), with a typical \textit{p}-value of 0.05, and 0.36 $\pm$ 0.04 in the right-hand panel (where def$_{\text{\HI}} \equiv$ def$_{\text{\HI, }M_\star}$), with a typical \textit{p}-value of 0.01.} Uncertainties on the Spearman correlation coefficient were obtained using a Monte Carlo analysis, and represent the spread in the values of Spearman's correlation coefficient after calculating it $1 \times 10^5$ times, \drafttwo{perturbing each point by its errorbar.} The weak correlation between \HI\ and H$_2$ deficiency suggests that the environmental effects removing \HI\ from galaxies also affect the molecular gas, albeit to a lesser extent.

Most galaxies are located in the upper-right quadrant of Figure \ref{fig:HI_H2_def}\draftthree{, meaning} they are both \HI\ and H$_2$ deficient. This suggests that, \draftthree{despite the lack of a stronger correlation between their deficiencies,} both the atomic and molecular gas phases are affected by environmental processes.

\begin{figure*}
	\centering
	\includegraphics[width=0.75\textwidth]{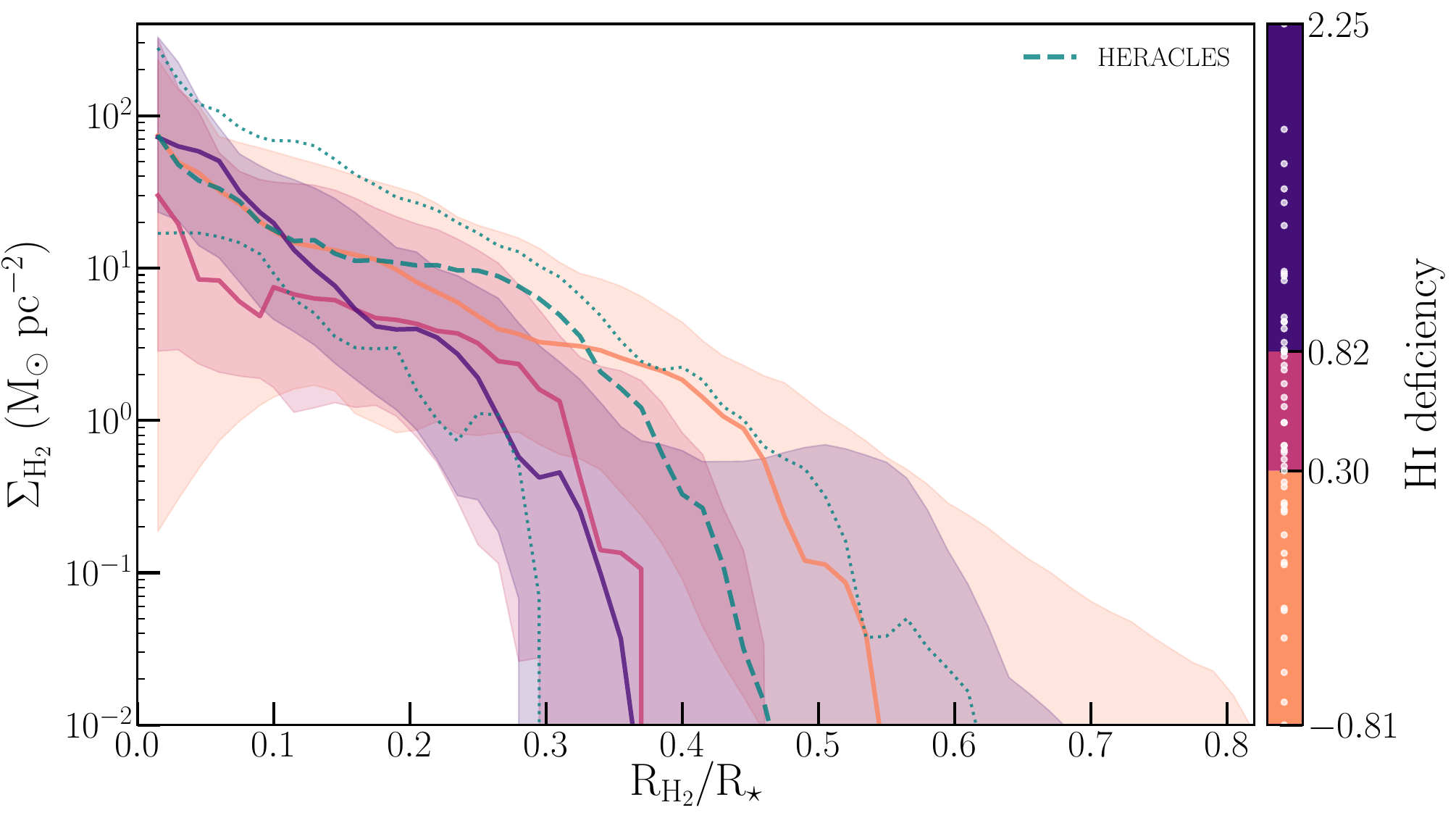}
	\caption{\draftfour{Observed median} H$_2$ radial profiles of VERTICO galaxies shown in three bins of $\sim$15 galaxies, sorted by \HI\ deficiency. The shaded areas highlight the 16\textsuperscript{th}-to-84\textsuperscript{th} percentile regions. The teal \draftthree{dashed} line represents the median profile of 21 HERACLES field galaxies. \draftthree{Its 16\textsuperscript{th}-to-84\textsuperscript{th} percentiles are indicated with dotted lines.} The distribution of the \HI\ deficiencies in the sample is shown on the colour bar using white dots. More \HI\ deficient galaxies have steeper and more truncated H$_2$ radial profiles. \draftthree{The median radial profile of HERACLES galaxies lies between those of the most \HI-rich and moderately \HI-deficient VERTICO galaxies.}}
	\label{fig:H2_rad_profs}
\end{figure*}

\drafttwo{In terms of \HI\ classes,} there is \drafttwo{a} significant scatter in H$_2$ deficiency within each class. There are, however, some differences between the classes. \drafttwo{Except \draftthree{for} a} few outliers, class \textsc{0} galaxies (galaxies with ``normal'' \HI\ reservoirs which resemble those of field galaxies) have normal \draftthree{to high} H$_2$ \draftthree{fractions}, in line with their morphologically normal, non-deficient \HI\ reservoirs. Class \textsc{i} galaxies, which exhibit one-sided \HI\ features, but otherwise have \HI\ and stellar discs similar to those of field galaxies, are H$_2$ normal. Classes \textsc{ii} and \textsc{iii} (galaxies with moderate to extreme asymmetries and significant \HI\ deficiencies) have moderately deficient H$_2$ reservoirs. \draftthree{Galaxies that are \HI\ deficient due to their lower \HI\ surface densities (class \textsc{iv} galaxies) have relatively high H$_2$ deficiencies.} \draftthree{The six known RPS galaxies are relatively H$_2$-rich compared to other VERTICO galaxies with similar \HI\ deficiencies.}  

\begin{figure*}[b!]
	\centering
	
	\subfloat
	{\includegraphics[width=0.43\textwidth]{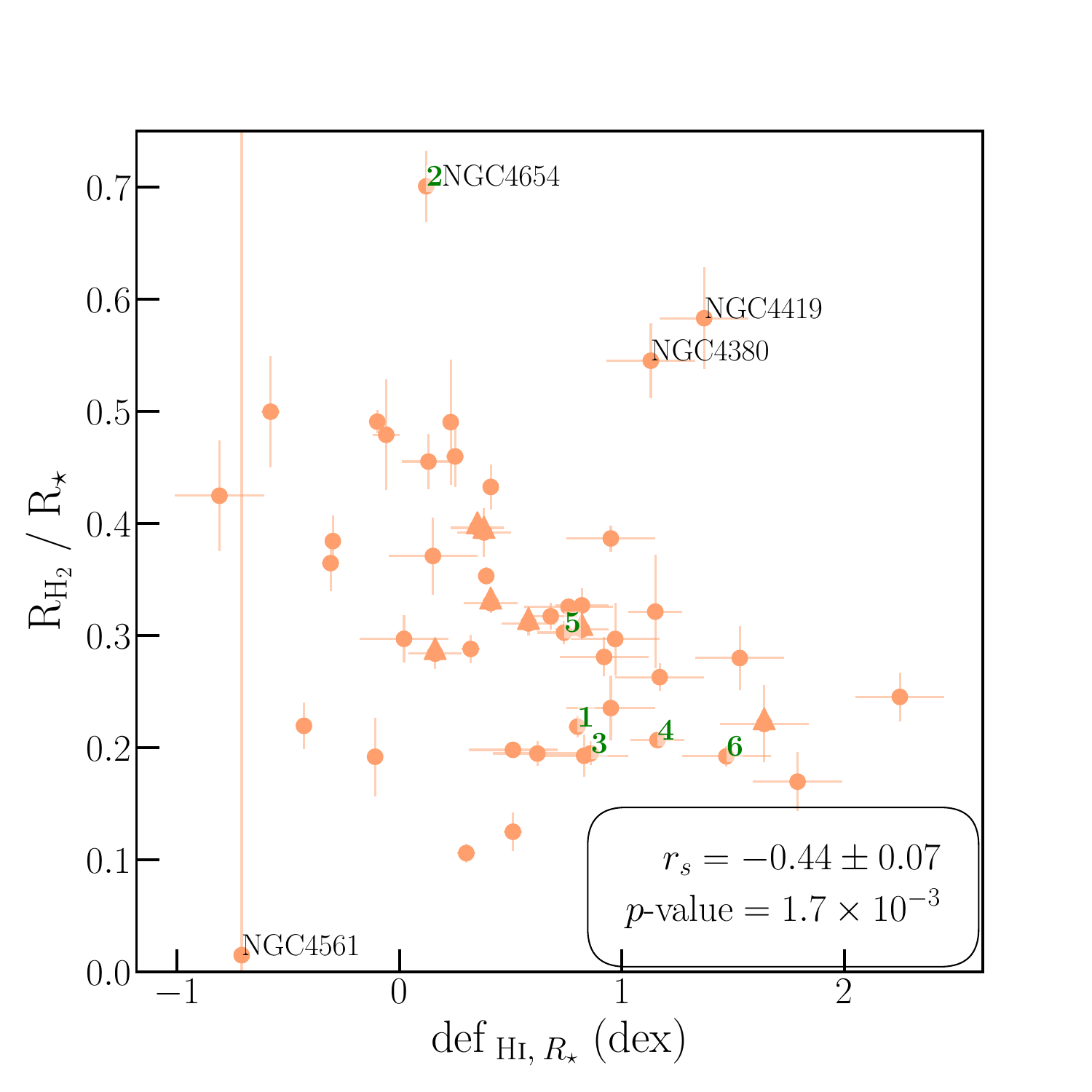}} \hspace{-8.5mm}
	\subfloat
	{\includegraphics[width=0.44\textwidth]{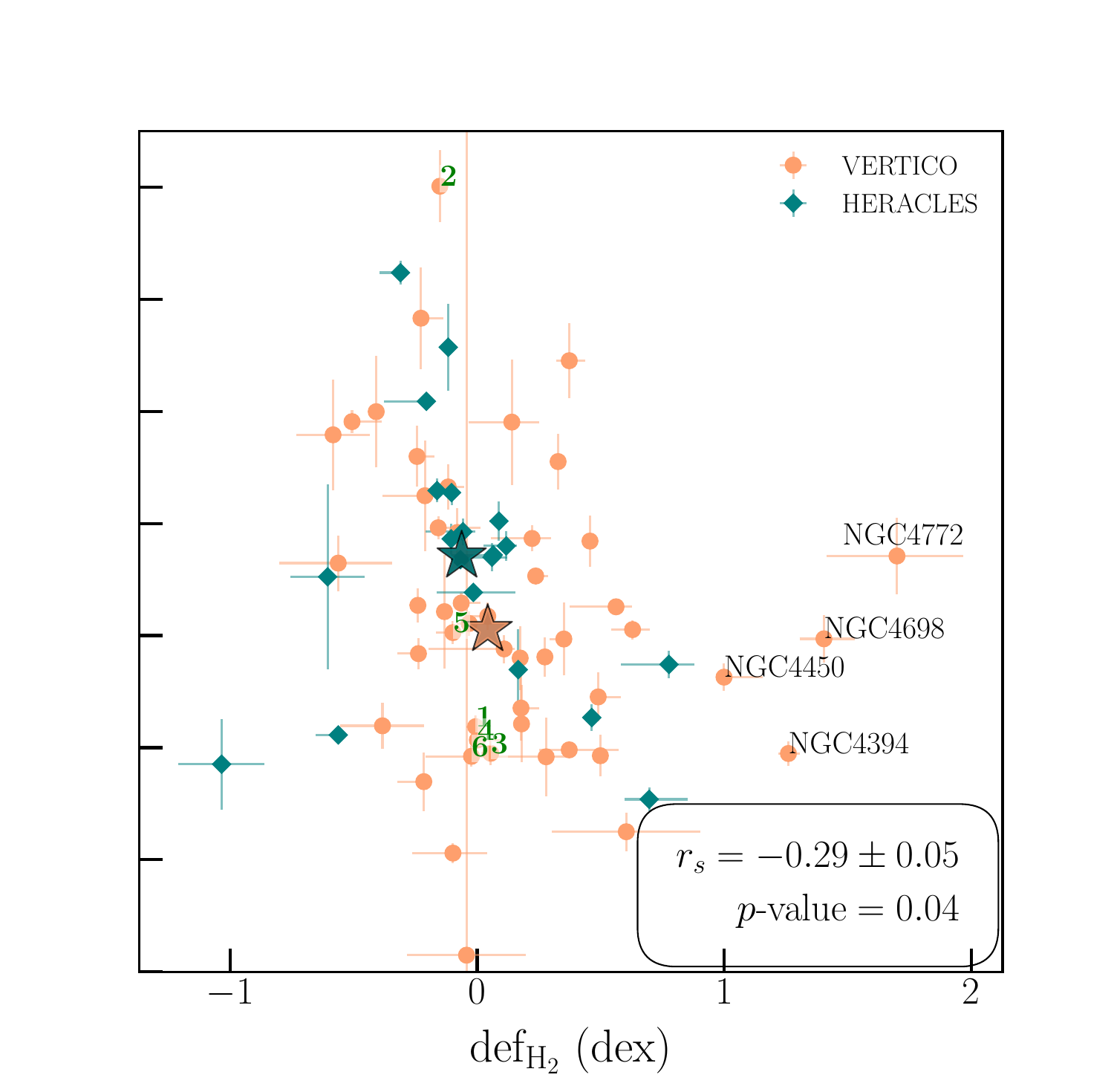}}
	
	\caption{Radius of the molecular gas disc (normalised by the radius of the stellar disc) as a function of \HI\ deficiency (left panel) and H$_2$ deficiency (right panel). Markers are the same as in Figure \ref{fig:H2_frac}, \draftfour{but with the numbers indicating RPS galaxies now located to the upper-right of the relevant markers}. \refrep{In the right-hand panel the median values (both in x and y) of the VERTICO and HERACLES samples are indicated with orange and teal stars, respectively.} The H$_2$ radius at an \draftfour{observed} isodensity threshold of $\Sigma_{\text{H}_2} = 1\ \text{M}_\odot\ \text{pc}^{-2}$ is \drafttwo{plotted}. There is an anti-correlation between \HI\ deficiency and $R_{\text{H}_2}$ (the Spearman correlation coefficient is $r_s = -0.44 \pm 0.07$ with a typical \textit{p}-value of \draftthree{$1.7 \times 10^{-3}$}). The relation between H$_2$ deficiency and $R_{\text{H}_2}$ at this isodensity threshold is \draftthree{weaker} (the Spearman correlation coefficient is $r_s = -0.29 \pm 0.05$, with a typical \textit{p}-value of 0.04). \HI\ deficiencies \drafttwo{are associated with} truncated H$_2$ discs, \draftthree{likely through outside-in stripping.}}
	\label{fig:H2_radii}
\end{figure*}

\begin{figure*}[b!]

	\centering
	
	\subfloat
	{\includegraphics[width=0.43\textwidth]{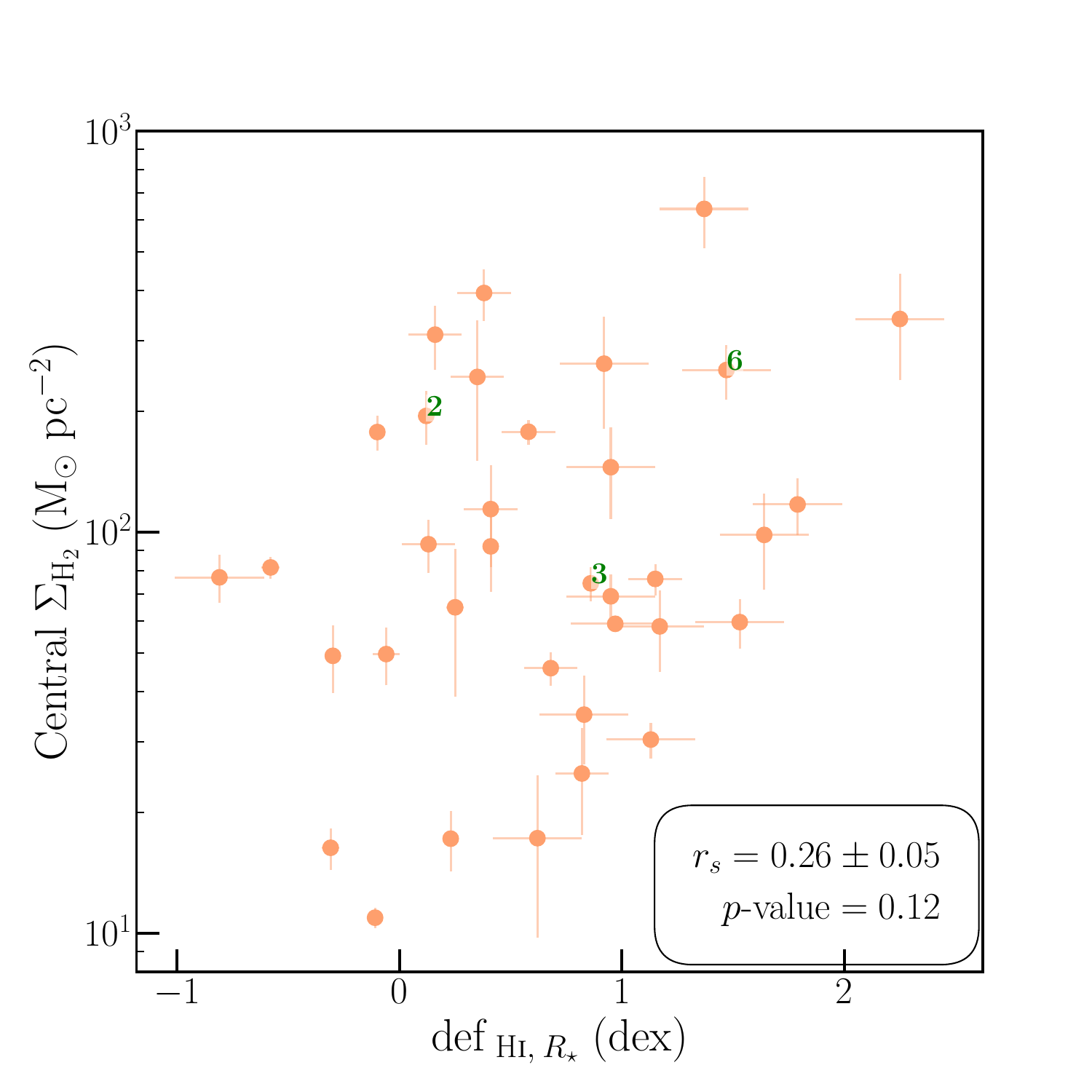}} \hspace{-8.5mm}
	\subfloat
	{\includegraphics[width=0.43\textwidth]{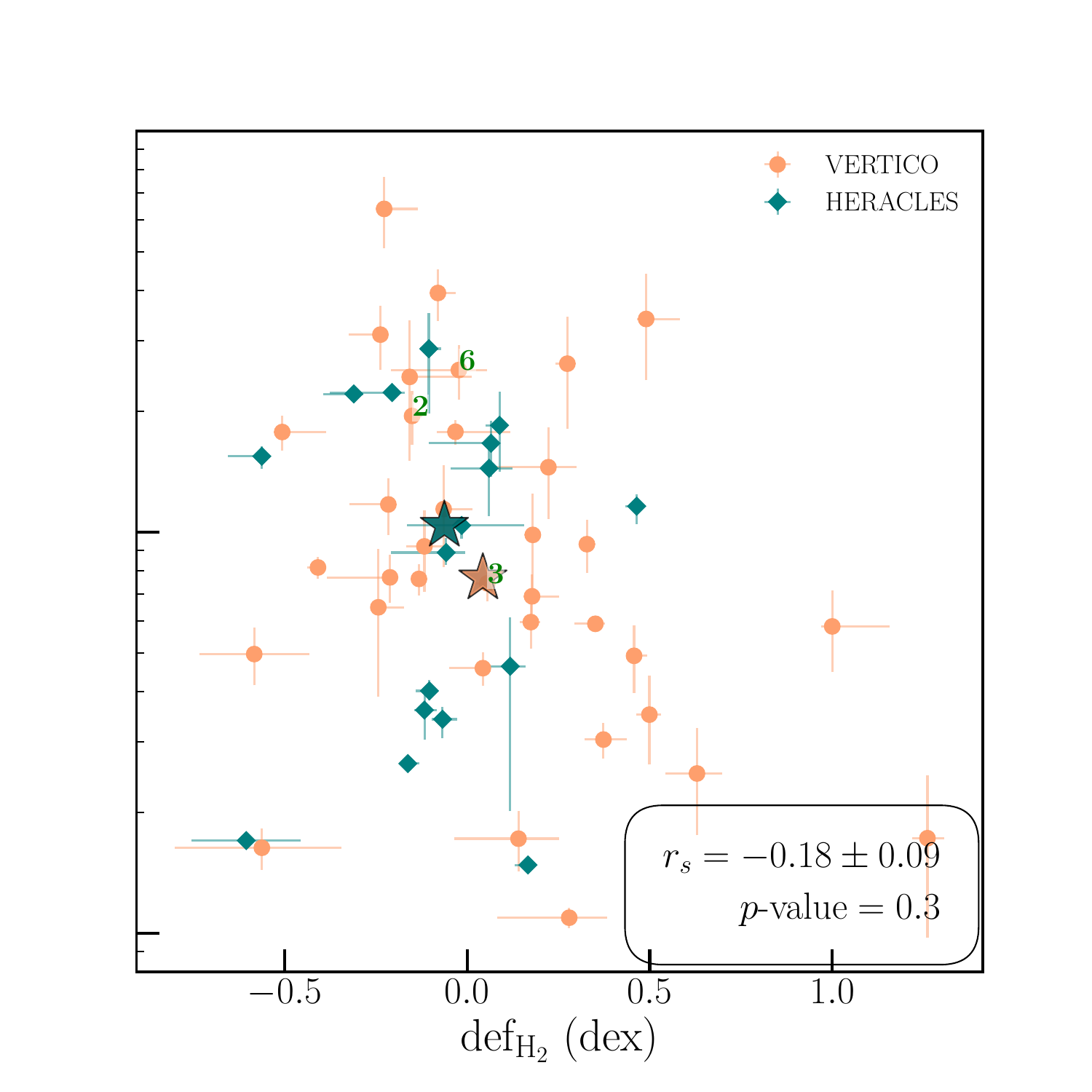}}
	
	\caption{Similar to Figure \ref{fig:H2_radii}, but showing median \draftfour{central} H$_2$ surface densities within \draftthree{an ellipse with \draftfour{an observed} semi-major axis of $0.1$R$_\star$} (e.g. the extreme centres of galaxies). \draftthree{While, by eye, there appears to be a weak correlation between \HI\ deficiency and $\Sigma_{\text{H}_2}$, it is not statistically significant (the Spearman correlation coefficient is $r_s = 0.26 \pm 0.05$, with a typical p-value of 0.12).} There is no correlation between H$_2$ deficiency and $\Sigma_{\text{H}_2}$ (the Spearman correlation coefficient is $r_s = -0.18 \pm 0.09$, with a typical p-value of 0.3). VERTICO galaxies have central median H$_2$ surface densities similar to those HERACLES galaxies \drafttwo{at \draftthree{similar} def$_{\text{H}_2}$}.}
	\label{fig:H2_densities}
\end{figure*}

\subsection{H$_2$ radial profiles by \HI\ deficiency} \label{sub:rad_prof_results}
\drafttwo{Median} $\Sigma_{\text{H}_2}$ radial profiles are shown in Figure \ref{fig:H2_rad_profs}\draftthree{, t}hey are divided into three equal bins of \drafttwo{def$_{\text{\HI,}\:R_\star}$}, indicated by the colour bar. In the colour bar, the white dots show the distribution of \HI\ deficiencies. The solid lines show the medians of the radial profiles in each bin; the shaded areas cover the 16\textsuperscript{th}-to-84\textsuperscript{th} percentile regions. The teal \draftthree{dashed} line represents the median H$_2$ radial profile of the 21 galaxies in the HERACLES field sample, \draftthree{whose 16\textsuperscript{th}-to-84\textsuperscript{th} percentiles are indicated with dotted lines}. The x-axis is normalised by the isodensity radius of the stellar disc \drafttwo{(at $\Sigma_{\star}(R_{\star})~=~1~\text{M}_{\odot}$~pc$^{-2}$}), the y-axis is not normalised. 

More strongly \HI-deficient galaxies have steeper and more truncated H$_2$ radial profiles. This suggests that the environmental processes acting on the atomic gas cause outside-in \drafttwo{removal} of the molecular gas. \draftthree{The median H$_2$ radial profile of the HERACLES sample lies between those of the most \HI-rich and moderately \HI\ deficient VERTICO galaxies.}

\subsection{H$_2$ Radii}
\draftone{Figure \ref{fig:H2_radii} shows the \drafttwo{H$_2$ disc radii} \draftfour{($R_{\text{H}_2}$)} as a function of \HI\ deficiency (left panel) and H$_2$ deficiency (right panel), for \draftthree{a $\Sigma_{\text{H}_2}$} threshold \draftthree{1} $\text{M}_\odot\ \text{pc}^{-2}$. As can be seen in Figure \ref{fig:H2_rad_profs}, this threshold \draftthree{probes} the low-density gas at the outskirts of the H$_2$ disc, which is \drafttwo{the gas} most likely to be affected by environmental mechanisms}. $R_{\text{H}_2}$ declines \draftthree{significantly} with \HI\ deficiency; the Spearman correlation coefficient is $r_s = -0.44 \pm 0.07$, \drafttwo{with a typical p-value of \draftthree{1.7 $\times 10^{-3}$. Outliers of this correlation are NGC4561, NGC4654, NGC4419, and NGC4380 (annotated in the Figure). NGC4561 only has a few ``blobs'' of detected molecular gas, which means that it does not have a well-defined H$_2$ radius. NGC4380 has a large, regular looking molecular gas disc. NGC4654 also has a large molecular gas disc, with a one-sided RPS feature. Although NGC4419 is not among the ``well-studied'' RPS cases, the morphology of its molecular gas disc suggests that it is likely also undergoing RPS (see Appendix A in B21).}}

\draftthree{There is a weaker anti-correlation between H$_2$ radius and H$_2$ deficiency, which is likely intrinsic. The Spearman correlation coefficient of this relation is $r_s = -0.29 \pm 0.05$, with a typical $p$-value of 0.04}. \draftthree{Outliers of this relation are NGC4772, NGC4698, NGC4450, and NGC4394. Like NGC4561, NGC4772 only has very little, patchy detected molecular gas, which results in a large radius for its strong H$_2$ deficiency. In NGC4698, the molecular gas is distributed in a ring, resulting in a similar effect. NGC4450 has a one-sided molecular gas feature, which is responsible for its relatively large H$_2$ radius. The molecular gas in NGC4394 is also patchy, again resulting in a strong deficiency compared to its H$_2$ radius.}

The anti-correlation between \HI\ deficiency and $R_{\text{H}_2}$ suggests that the environmental mechanism(s) removing \HI\ from galaxies also cause outside-in stripping of their molecular gas. \refrep{On average, galaxies in the HERACLES sample have marginally smaller H$_2$ deficiencies and marginally larger H$_2$ radii. However, there is no statistically significant difference between both samples.}

\begin{figure*}
	\centering
	
	\subfloat
	{\raisebox{-0.8mm}{\includegraphics[width=0.5\textwidth]{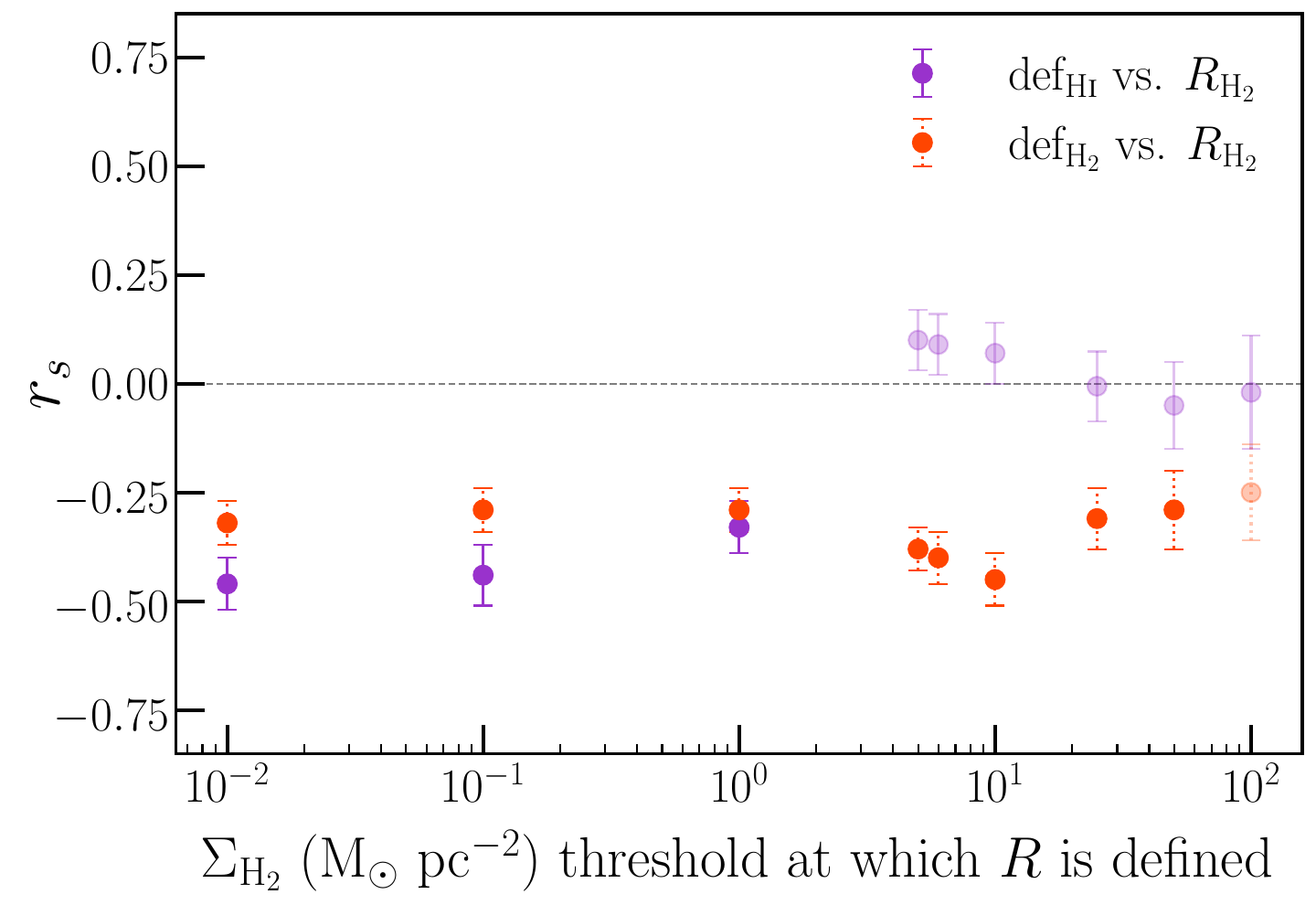}}}
	\subfloat
	{\includegraphics[width=0.45\textwidth, height=0.340\textwidth]{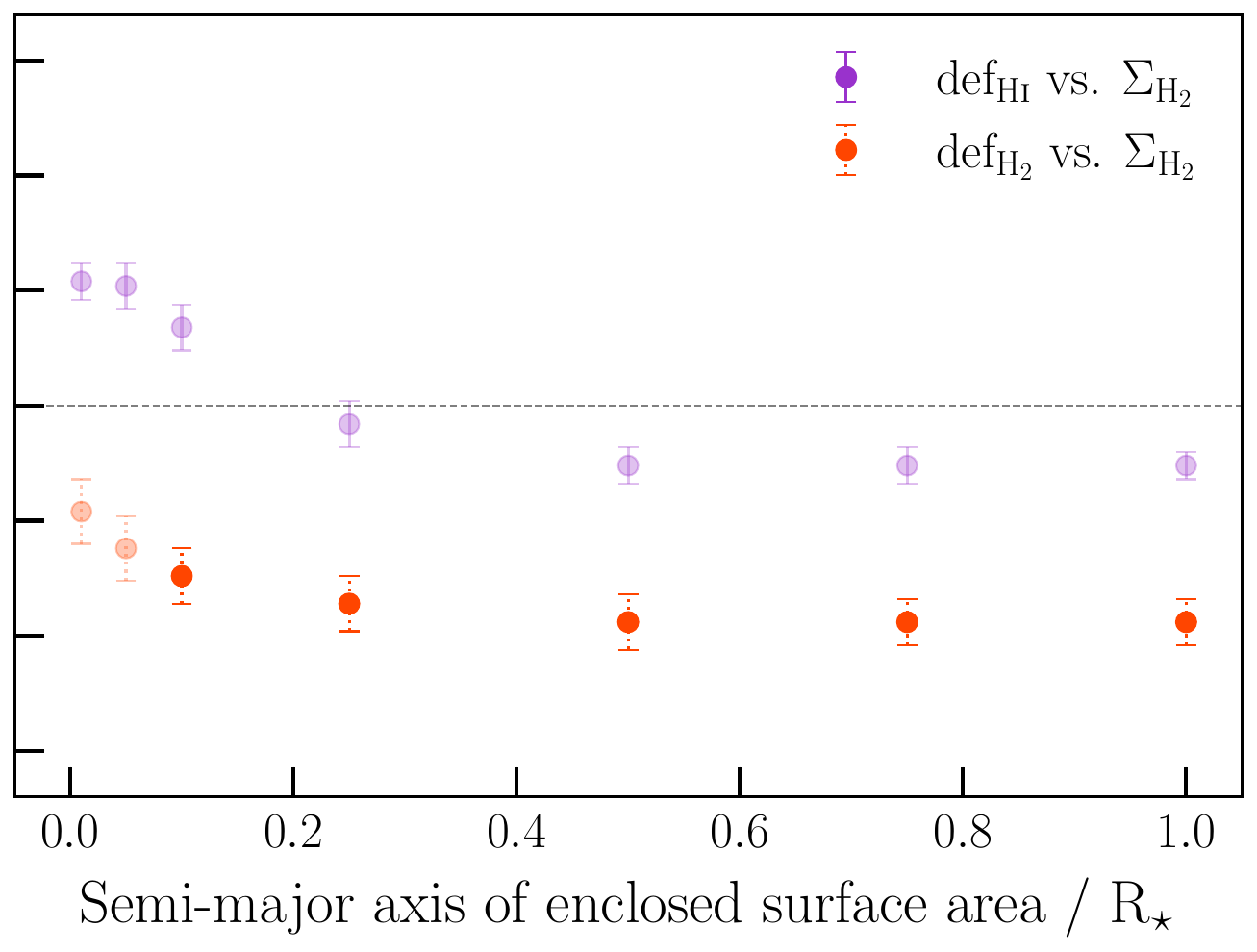}}	
	
	\caption{\draftone{\emph{Left panel:} Spearman correlation coefficients as a function of \draftthree{the \draftfour{observed} isodensity threshold at which $R_{\text{H}_2}$ is defined for the relationship between \HI\ (purple) and H$_2$ (orange) deficiency and $R_{\text{H}_2}$}. \draftthree{Spearman test results with $p$-values greater than 0.05 (i.e. relationships with no significant correlation) are indicated with semi-transparent markers. There is an anti-correlation between $R_{\text{H}_2}$ and both \HI\ and H$_2$ deficiency. The anti-correlation between $R_{\text{H}_2}$ and \HI\ deficiency only exists if the isodensity radius of the H$_2$ disc is defined at low H$_2$ surface densities, i.e. the very outskirts of the disc. This is consistent with an outside-in stripping scenario of H$_2$ in \HI\ deficient galaxies.} \newline \emph{Right panel:} Spearman correlation coefficients as a function of \draftthree{the \draftfour{observed} semi-major axis of the ellipse in which the median H$_2$ surface density is calculated, for the relationships between \HI\ (purple) and H$_2$ (orange) deficiency and H$_2$ surface density. There is an anti-correlation between H$_2$ deficiency and $\Sigma_{\text{H}_2}$, which is likely intrinsic. Although there is a hint of slightly increased central H$_2$ surface densities, the relationship between \HI\ deficiency and $\Sigma_{\text{H}_2}$ is not statistically significant.}}}
	\label{fig:spearman_radius}
\end{figure*}

\subsection{\draftfour{Central} H$_2$ surface densities} \label{sub:h2_surf_dens}
Figure \ref{fig:H2_densities} shows the median surface density of the molecular gas disc as a function of \HI\ deficiency (left panel) and H$_2$ deficiency (right panel), \draftone{within a radius of \draftthree{$R = 0.1\: \text{R}_\star$} \drafttwo{(typically corresponding to several \draftthree{hundred} pc, roughly equivalent to the inner resolution element of the moment 0 map for most galaxies in the sample)}. \draftfour{This radius was chosen because, if} any \drafttwo{inward} compression of the gas is taking place, the effect is likely subtle and primarily visible in the very centre of the galaxy.} \draftthree{While, by eye, there appears to be a weak correlation between \HI\ deficiency and central H$_2$ surface density, it is not statistically significant (the Spearman correlation coefficient is $r_s = 0.26 \pm 0.05$ with a typical p-value of 0.12)}.

There is no correlation between H$_2$ deficiency and \draftfour{central} H$_2$ surface density (the Spearman correlation coefficient is $r_s = -0.18 \pm 0.09$, with a typical p-value of 0.3). \refrep{On average, galaxies in the HERACLES sample have marginally smaller H$_2$ deficiencies and marginally larger central H$_2$ surface densities. However, there is no statistically significant difference between both samples.}

\subsection{Effects of definition of H$_2$ radius} \label{sub:spearman}
Figures \ref{fig:H2_radii} and \ref{fig:H2_densities} are shown for specific choices of the H$_2$ radius: in Figure \ref{fig:H2_radii} the radius of the H$_2$ disc is defined at $\Sigma_{\text{H}_2} = 1\ \text{M}_\odot \text{ pc}^{-2}$ and in Figure \ref{fig:H2_densities} the median H$_2$ surface density is measured inside 0.1 R$_\star$. To investigate how the relationships shown in these plots change with the choice of radius \draftthree{(i.e. the isodensity threshold at which the H$_2$ radius is defined, and the semi-major axis of the ellipse enclosing the area in which the median H$_2$ surface density is calculated)}, we remake both \draftfour{Figures} using a number of different radii, and \drafttwo{measure} their Spearman correlation coefficients. Figure \ref{fig:spearman_radius} shows these Spearman correlation coefficients for the relationships between \HI\ deficiency (\draftfour{left-hand} panel) and $R_{\text{H}_2}$ and $\Sigma_{\text{H}_2}$ (purple and orange, respectively), and, similarly, the relationships between H$_2$ deficiency and $R_{\text{H}_2}$ and $\Sigma_{\text{H}_2}$ (\draftfour{right-hand} panel). \draftthree{Relationships for which the $p$-value associated with Spearman's correlation is greater than 0.05 \draftfour{(and are thus considered to not be statistically significant)} are indicated with semi-transparent markers.}

\draftone{The relationship between \HI\ deficiency and $R_{\text{H}_2}$ is strongest for low isodensity thresholds, and disappears for higher thresholds. \draftthree{Thus,} this \draftthree{correlation} is only seen when the lower density material at the outskirts of the molecular gas disc is taken into account. The higher density material at smaller radii is not affected significantly. This implies either outside-in stripping, or the removal of molecular gas throughout the disc, resulting in the low-surface density gas at the outskirts \draftthree{dropping} below our detection threshold.} \drafttwo{However, the latter explanation would result in H$_2$ radial profiles with shapes similar to those of field galaxies, but scaled down, which is not observed for VERTICO galaxies (see Figure \ref{fig:H2_rad_profs})}. \drafttwo{The relationship between H$_2$ deficiency and $R_{\text{H}_2}$ is present for all values of the isodensity threshold \draftthree{except the very highest}, and is likely intrinsic.}

\draftthree{The relationship between \HI\ deficiency and $\Sigma_{\text{H}_2}$ is not statistically significant, and any hint of increased H$_2$ surface densities is only present in the very centres of galaxies ($R < 0.1 \text{R}_\star$).} \drafttwo{The moderate negative correlation between H$_2$ deficiency and median H$_2$ surface density is independent of the choice of radius, although it is no longer statistically significant when the median H$_2$ surface density is calculated in the central regions only. This correlation is likely intrinsic.}

\section{Discussion} \label{sec:discussion}
\draftthree{Globally, while \HI\ deficiency only correlates weakly to moderately with H$_2$ deficiency, \HI\ deficient galaxies are often also H$_2$ deficient to varying degrees. This suggests that the environmental mechanisms acting on the \HI\ in Virgo cluster galaxies simultaneously affect their H$_2$. However, the lack of a stronger correlation between both deficiencies, in combination with smaller H$_2$ deficiencies compared to \HI\ deficiencies, implies that relatively little molecular gas has been removed from VERTICO galaxies. It is possible that H$_2$ deficiencies become more significant only after most of the \HI\ is removed.}

A closer look at the resolved properties of the molecular gas in these galaxies has revealed that absence of a strong correlation does not necessarily mean that the H$_2$ content of \HI\ deficient galaxies remains undisturbed by environmental process(es). $\Sigma_{\text{H}_2}$ radial profiles of galaxies with larger \HI\ deficiencies are more truncated and steeper than those of their more \HI\ rich counterparts, indicating outside-in removal of molecular gas. We explored this result in more detail by studying the radius and average surface density of the molecular gas disc as a function of both \HI\ and H$_2$ deficiency. This has confirmed the presence of an anti-correlation between \HI\ deficiency and the radius of the H$_2$ disc \drafttwo{(Figure \ref{fig:H2_radii})}. This correlation is only significant when the low-density molecular gas at the outskirts of the disc is taken into account (Figure \ref{fig:spearman_radius}). Furthermore, there is a hint of a slight increase in median central surface density ($R \lesssim 0.1 \text{R}_\star$) of the H$_2$ disc with increasing \HI\ deficiency, although this is not statistically significant. \draftfour{In this work we have opted not to correct for inclination (see \S \ref{sec:methods}). The application of inclination corrections does not significantly alter the trends in any of the Figures, and therefore the broad conclusions of the paper remain the same with or without these corrections.}

\draftthree{Our results partly agree with the} results from \citet{Mok2017}, who report steeper H$_2$ radial profiles of Virgo galaxies compared to those of galaxies in the field. \draftthree{They also find enhanced central surface densities, for which we do not find solid evidence here}. \citet{Mok2017} mainly attribute this enhancement to radially inward migration of \HI\ as a result of RPS, after which it is more easily converted to H$_2$. They also report increased global molecular gas masses of Virgo galaxies compared to field and group samples, which \draftthree{while here we mostly find H$_2$ deficiencies compared to \draftthree{main sequence galaxies from} xCOLD GASS.} \draftthree{This difference is possibly the result of the use of a \HI-mass selected sample by \citet{Mok2017}.}

Our results also agree partly with those from \citet{Boselli2014}. While they indeed find \drafttwo{an anti-correlation} between H$_2$ radius and \HI\ deficiency, \draftthree{they also report a strong correlation between \HI\ and H$_2$ deficiency, whereas we find a weak correlation. The strong correlation between \HI\ and H$_2$ deficiency found by \citet{Boselli2014} is based on a sub-sample of 17 galaxies with high-quality data from  \citet{Kuno2007}. When the entire Virgo sample is taken into account, this relationship is not as strong (see figure 5 in \citealt{Boselli2014}). The range of \HI\ deficiencies considered in this work is closer to that of the entire sample of \citet{Boselli2014}, rather than the sub-sample from \citet{Kuno2007}. This sub-sample exclusively contains galaxies with \HI\ deficiencies between $\sim$0 and 1, where the slope is steepest (Figure \ref{fig:HI_H2_def}). Therefore, it is possible that the relationship found in this work is weaker because of the wider range of \HI\ deficiencies, similar to figure 5 of \citet{Boselli2014}.} \draftthree{Uncertainties in X\textsubscript{CO} and $R_{21} \equiv$ CO(2\draftthree{--}1)/CO(1\draftthree{--}0) may contribute to differences with previous work, although \citet{Boselli2014} report a stronger correlation between \HI\ and H$_2$ deficiency for a constant X\textsubscript{CO}.}

\subsection{What causes \HI\ and H$_2$ deficiencies?}
\draftthree{Truncated molecular gas discs could be explained by several environmental mechanisms}, such as RPS, tidal interactions, and galaxy-galaxy encounters. \draftfour{However, tidal interactions and galaxy-galaxy encounters are expected to result in significant kinematic in addition to morphological disturbances}, which are not \drafttwo{evident} in the moment 1 maps of VERTICO galaxies (see Appendix A in B21). Thermal evaporation can remove gas from the disc, \draftthree{which could result in observed truncation as the surface density of the molecular gas at the outskirts of the disc drops below the sensitivity limit of our observations}. The same is true for starvation. \draftthree{However, both these processes should result in relatively uniform removal of gas, which should not significantly change the shape of the $\Sigma_{\text{H}_2}$ radial profile.} Additionally, the large amounts of hot ICM required for thermal evaporation would automatically result in RPS as satellite galaxies move through it. 

\draftthree{The most likely explanation for the observed truncated molecular gas discs is RPS.} In the earlier stages of RPS, the outer parts of the radial profile are likely dominated by the gas tail, resulting in a relatively flat and extended \drafttwo{radial} profile. It is possible that, as the stripping continues, the outer parts of the stripped tail become undetectable, resulting in a steeper and more truncated profile. \draftthree{Additionally, RPS could result in the inward compression of gas, which would contribute to the steepness of the $\Sigma_{\text{H}_2}$ radial profiles, as \draftfour{also} reported by e.g. \citet{Mok2017}.} 

\draftthree{Several moment zero maps of VERTICO galaxies exhibit clear RPS signatures (for example NGC4654; B21, figure 4.22).} If RPS is responsible for the truncation of the molecular gas discs, the lack of a \draftthree{stronger} correlation between \HI\ and H$_2$ deficiency suggests that it is not very effective at removing molecular gas from galaxies, \draftthree{or at least not in the early stages, when significant amounts of \HI\ are still present}. This is also reflected in Figure \ref{fig:HI_H2_def}, which shows that the Virgo spirals with the clearest RPS signatures (including post-RPS peak galaxies) have relatively large H$_2$ fractions compared to other VERTICO galaxies with similar stellar masses. It is possible that only a small amount of low-density gas is stripped \draftthree{(at this stage)}. 

\refrep{One of the VERTICO galaxies with clear RPS signatures, NGC4402, has recently been studied in detail by \citet{Cramer2020} using high-resolution ($1^{\prime\prime} \times 2^{\prime\prime}$, $\sim$80--160 pc at the distance of the Virgo cluster) CO(2~--~1) observations from ALMA. They estimate the strength of the ram pressure at the location of NGC4402, and find that only diffuse, low-surface density H$_2$ (up to $2.5-10~\text{M}_\odot\ \text{pc}^{-2}$, depending on the exact ICM density at its location) can be stripped from this galaxy by RP. This limit is likely similar or lower for other galaxies in the VERTICO sample, as NGC4402 is probably close to the cluster centre ($\sim$0.4 Mpc in projection), and ram pressure decreases rapidly with cluster-centric radius ($p \propto r^{-2}$, \citealt{Gunn1972}). This is in line with our findings, which suggest that RPS is acting on the molecular gas in VERTICO galaxies, but unable to remove significant amounts of it.}

\refrep{\citet{Cramer2020} also report a region with increased CO surface brightness as a result of RPS in NGC4402, resulting in enhanced star formation (which aids the removal of diffuse gas through stellar feedback). In a similar study of the CO in Coma cluster galaxy NGC 4921, \citet{Cramer2021} also find evidence for compression of the dense ISM on the leading side of this galaxy, accompanied by enhanced star formation activity. As discussed above, similar results were found by \citet{Mok2017}. Such increased molecular gas densities could explain the relatively large molecular gas fractions in well-studied RPS galaxies in the VERTICO sample (Figure \ref{fig:HI_H2_def}), by offsetting the amount of diffuse H$_2$ lost through stripping. We do not find evidence for increased H$_2$ surface densities in this work. However, here we exclusively work with radially averaged radial profiles. More detailed (case) studies of the H$_2$ distribution and SFRs in VERTICO galaxies are needed to investigate how common regions with increased H$_2$ surface densities and star formation as a result of RPS are, and to what extent (and on what timescales) they can compensate the mass loss from the stripping of low-surface density molecular gas.}

Galaxies that do not show any sign of environmental effects on their \HI\ reservoirs (class \textsc{0} galaxies in \citealt{Yoon2017}) are H$_2$-rich compared to galaxies that do. \draftfour{The difference in H$_2$ reservoirs between \HI-normal and \HI-disturbed objects confirms that the effects of environment on atomic gas also act on molecular gas.} Galaxies with relatively low \HI\ surface densities (class \textsc{iv} galaxies) are relatively H$_2$-poor. This suggests that (external or internal) processes resulting in lower gas surface densities are more effective at reducing the molecular gas reservoir than \draftthree{outside-in stripping}. Galaxies with very strongly asymmetric and/or truncated \HI\ discs cover a range in H$_2$ deficiencies, suggesting that the mechanisms causing these features do not necessarily significantly reduce the global molecular gas mass, \draftthree{at least on timescales where the galaxies still contain detectable \HI.}

\draftthree{Simulations of RPS acting on the multi-phase ISM (with H$_2$ treated as a separate component) have indeed shown that RPS can strip H$_2$ from the outer parts of galaxies, especially when the orientation of the wind is face-on \citep{Lee2020}. However, not all simulations agree on how molecular gas in impacted by RPS. \citet{Tonnesen2009} have shown, for example, that the fate of dense gas depends heavily on the strength of the RP: at low RP the amount of high density gas is enhanced, while it is depressed at high RPs. Similarly, \citet{Tonnesen2019} find that in recent infallers compression can lead to high density gas, that is impossible to be stripped at later stages. Thus, whether significant amounts of H$_2$ are stripped, and whether H$_2$ is enhanced in galactic centres, likely varies between clusters, and depends on the locations of galaxies in phase-space.}

\draftfour{While simulations have rarely mapped H$_2$,} some simulations show mild SF enhancement in RPS galaxies, particularly at the pericentre of their orbits \citep{Bekki2014, Steinhauser2016, Ruggiero2017}. However, not all of them agree on where in galaxies these enhancements are found. \citet{Tonnesen2012}, \citet{Bekki2014}, and \citet{Boselli2021} find that SF in generally enhanced in the centres, while \citet{Roediger2014} find that SF is only enhanced via compression at the disk edges, right before gas is removed. All in all, there is little evidence for central enhancement of SF from simulations, and therefore, assuming that SF traces H$_2$, of central enhancement of H$_2$.

\section{Summary} \label{sec:summary}
In this work we have investigated how environmental mechanisms acting on atomic gas affect the molecular gas in spiral galaxies in \draftthree{the Virgo cluster}. Observations of 49 spiral galaxies in the Virgo cluster from the ALMA large program VERTICO: ``the Virgo Environment Traced in CO'' were used to study the resolved properties of the molecular gas. Their \HI\ deficiencies, as well as deficiency/morphology classifications, were adopted from the VLA Imaging of Virgo in Atomic gas survey (VIVA, \citealt{Chung2009, Yoon2017}). In particular, we studied how the H$_2$ radial profile, radius, and surface density change as a function of \HI\ deficiency, H$_2$ deficiency, and between different \HI\ deficiency/morphology classes. Our main findings are as follows:
\begin{itemize}
\item \refrep{Many VERTICO galaxies that are \HI\ deficient are also somewhat H$_2$ deficient, albeit to a lesser extent.} There is a weak correlation between \HI\ deficiency and H$_2$ deficiency, although it exhibits significant scatter. This suggests that, although environmental effects simultaneously affect H$_2$ and \HI\ in VERTICO galaxies, \HI\ deficiency does not always predict H$_2$ deficiency (i.e. when significant amounts of \HI\ are removed from galaxies, they often still contain significant amounts of H$_2$).
\item \drafttwo{Galaxies with larger \HI\ deficiencies} have steeper and less extended molecular gas radial profiles. This means that the mechanism removing atomic gas from galaxies causes truncation of the molecular gas disc.
\item \drafttwo{We find an anti-correlation between \HI\ deficiency and H$_2$ radius, which confirms that the environmental mechanisms removing atomic gas from galaxies remove or redistribute molecular gas from the outside-in. This correlation is only significant when radius is measured at the outskirts of the H$_2$ disc, which implies that only the low-density gas at large radii is affected.}
\item \draftthree{The most likely explanation for the observed steepened shapes of the $\Sigma_{\text{H}_2}$ radial profiles of \HI\ deficient VERTICO galaxies is RPS, as other environmental processes are expected to result in the more uniform removal of H$_2$.}
\item Galaxies that show clear signs of ongoing ram pressure stripping are H$_2$-normal to H$_2$-rich. \draftthree{This implies that RPS is not effective at reducing global molecular gas fractions on the timescales in which such features are still clearly visible.}
\item \draftthree{Galaxies with low \HI\ surface densities rather than asymmetries or truncation (class \textsc{iv} galaxies) are relatively H$_2$-poor. This suggests that additional external or internal mechanisms which reduce the gas density are more effective at reducing H$_2$ fractions than external processes causing outside-in stripping (i.e. RPS).}
\end{itemize}

\draftthree{In summary, the atomic and molecular gas in VERTICO galaxies are affected by environment simultaneously. While \HI\ deficiency is not a very accurate predictor of global H$_2$ deficiency, \HI\ deficient VERTICO galaxies show clear signs of outside-in removal of molecular gas.}

\section{Acknowledgements} \label{sec:acknowledgements}
This work was carried out as part of the VERTICO collaboration. 

The authors thank Cecilia Bacchini for pointing out errors in Table \ref{tab:VERTICO-sample}, which have been corrected in this version of the paper.

\draftfour{CDW acknowledges support from the Natural Sciences and Engineering Research Council of Canada and the Canada Research Chairs program.}

\draftthree{BL acknowledges support from the National Science Foundation of China (12073002, 11721303).}

\draftthree{ARHS is a grateful recipient of the Jim Buckee Fellowship at ICRAR.}

IDR acknowledges support from the ERC Starting Grant Cluster Web 804208.

AC acknowledges the support from the National Research Foundation grant No. 2018R1D1A1B07048314.

\draftfour{KS and LCP acknowledge support from the Natural Sciences and Engineering Research Council of Canada.}

\draftfour{ST acknowledges support from the Simons Foundation.}

\draftthree{LC acknowledges support from the Australian Research Council Discovery Project and Future Fellowship funding schemes (DP210100337, FT180100066). Parts of this research were conducted by the Australian Research Council Centre of Excellence for All Sky Astrophysics in 3 Dimensions (ASTRO 3D), through project number CE170100013.}

\draftfour{VV} acknowledges support from the scholarship ANID-FULBRIGHT BIO 2016-56160020, 
funding from NRAO Student Observing Support (SOS)-SOSPA7-014, 
and from partial support from NSF-AST1615960. 

\draftthree{Parts of this research were supported by the Australian Research Council Centre of Excellence for All Sky Astrophysics in 3 Dimensions (ASTRO 3D), through project number CE170100013.}

This work made use of HERACLES, `The HERA CO-Line Extragalactic Survey’ \citep{Leroy2009}.

This paper makes use of the following ALMA data: ADS/JAO.ALMA\# 2019.1.00763.L, \newline ADS/JAO.ALMA\# 2017.1.00886.L, \newline ADS/JAO.ALMA\# 2016.1.00912.S, \newline ADS/JAO.ALMA\# 2015.1.00956.S. \newline ALMA is a partnership of ESO (representing its member states), NSF (USA) and NINS (Japan), together with NRC (Canada), MOST and ASIAA (Taiwan), and KASI (Republic of Korea), in cooperation with the Republic of Chile. The Joint ALMA Observatory is operated by ESO, AUI/NRAO and NAOJ. In addition, publications from NA authors must include the standard NRAO acknowledgement: The National Radio Astronomy Observatory is a facility of the National Science Foundation operated under cooperative agreement by Associated Universities, Inc.

\software{\texttt{Matplotlib} \citep{Hunter2007}, \texttt{NumPy} \citep{Harris2020}, \texttt{Astropy} \citep{Astropy2013, Astropy2018}, \texttt{SciPy} \citep{2020SciPy}, \texttt{Photutils} \citep{Bradley2020}, and \texttt{Lifelines} \citep{Davidson-Pilon2019}}.

\clearpage

\appendix
\renewcommand\thefigure{\thesection.\arabic{figure}}

\section{Additional figures} \label{app:plots}
\setcounter{figure}{0}   
\draftthree{This Appendix contains additional Figures that are referred to at various points throughout the paper.}

\begin{figure}
	\centering
	\includegraphics[width=0.9\textwidth]{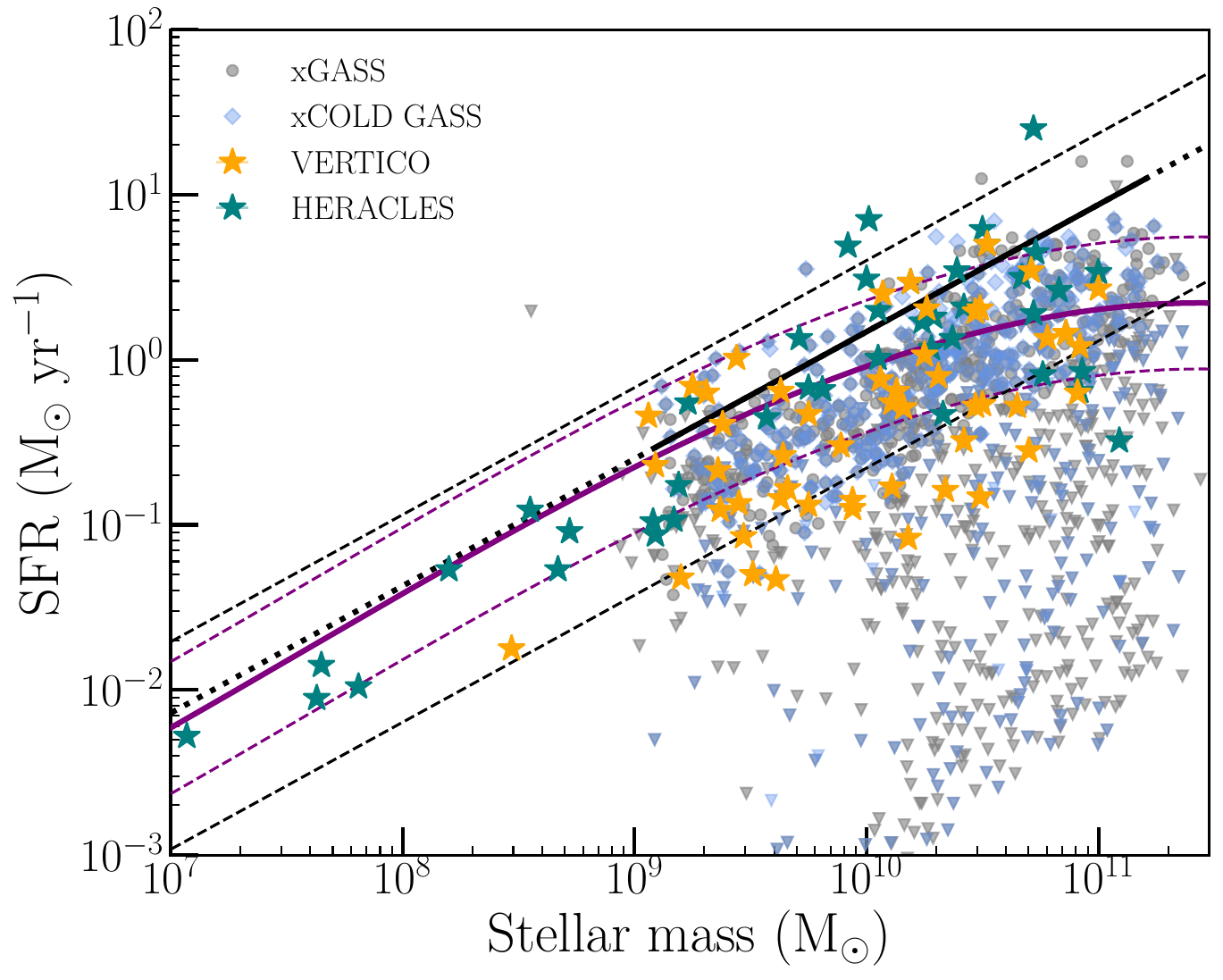}
	\caption{\drafttwo{SFR-$M_\star$ plot for VERTICO (orange) and the control samples used in this work (xGASS in grey, xCOLD GASS in blue, and HERACLES in teal). Overlaid are the SFMS from \citet{Elbaz2007} in black (the thick, dotted line is an extrapolation of the solid line covering the stellar mass range in which the relationship was derived), and that from xCOLD GASS in purple \citep{Saintonge2016}. \draftfour{Stellar masses and star formation rates for the x(COLD) GASS samples were adopted from GSWLC for consistency with VERTICO, for which they are adopted from z0MGS. Objects for which no 22 $\mu$m detection is available are indicated with down-pointing triangles.} Following the style of \citet{Saintonge2016}, errorbars are omitted to improve readability. For the control sample used in this work, xGASS and xCOLD GASS galaxies within 2$\sigma$ of the SFMS from \citet{Elbaz2007} are used, as indicated by the black dashed lines. While VERTICO and x(COLD) GASS are primarily scattered below the SFMS, HERACLES galaxies are distributed around higher SFRs. This is likely a result of selection.}}
	\label{fig:SFMS}
\end{figure}

\begin{figure}
	\centering
	
	\subfloat
	{\includegraphics[width=0.48\textwidth]{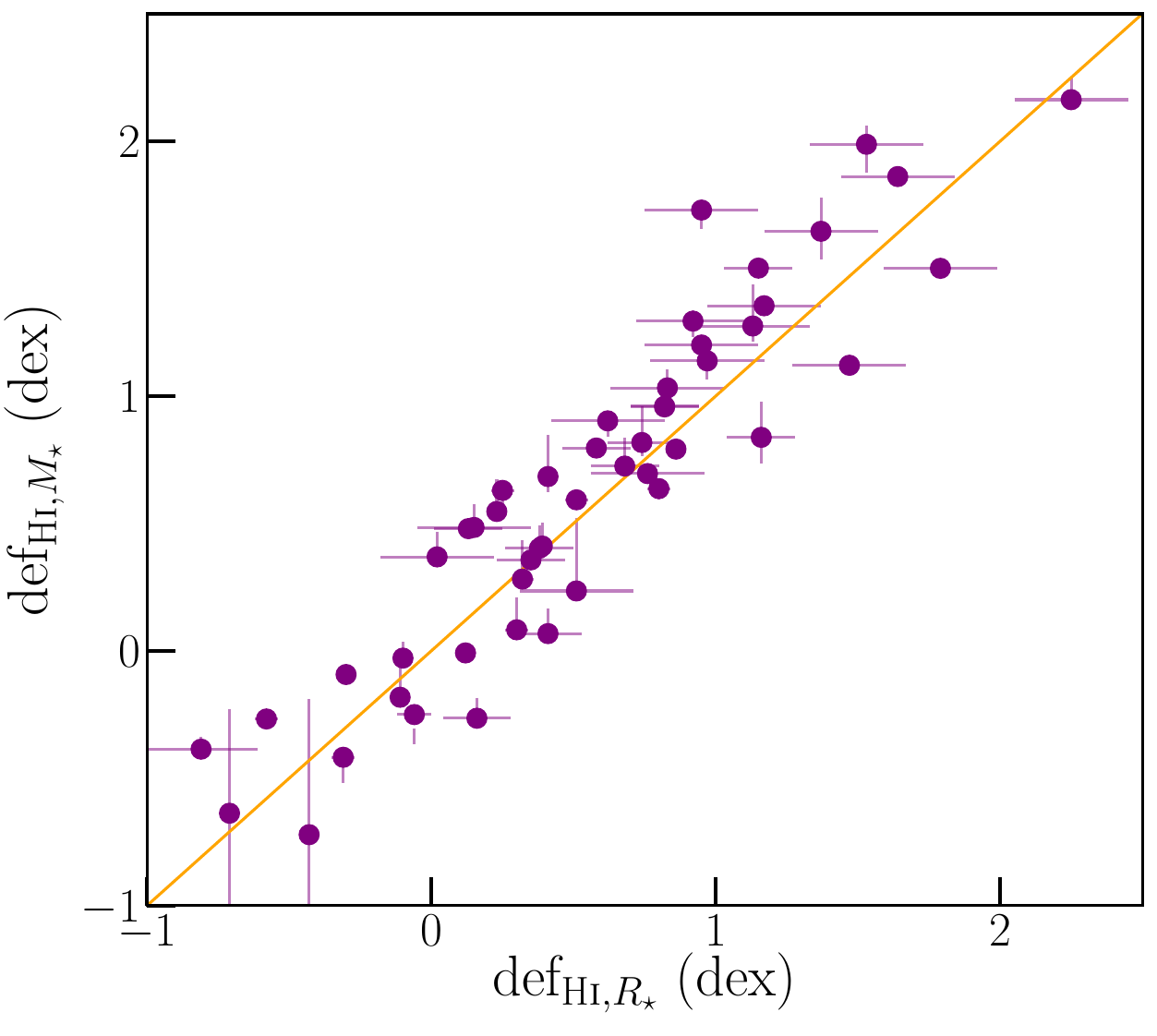}\label{fig:hi_def_vs_hi_def}}
	\hspace{0.5cm}
	\subfloat
	{\includegraphics[width=0.48\textwidth]{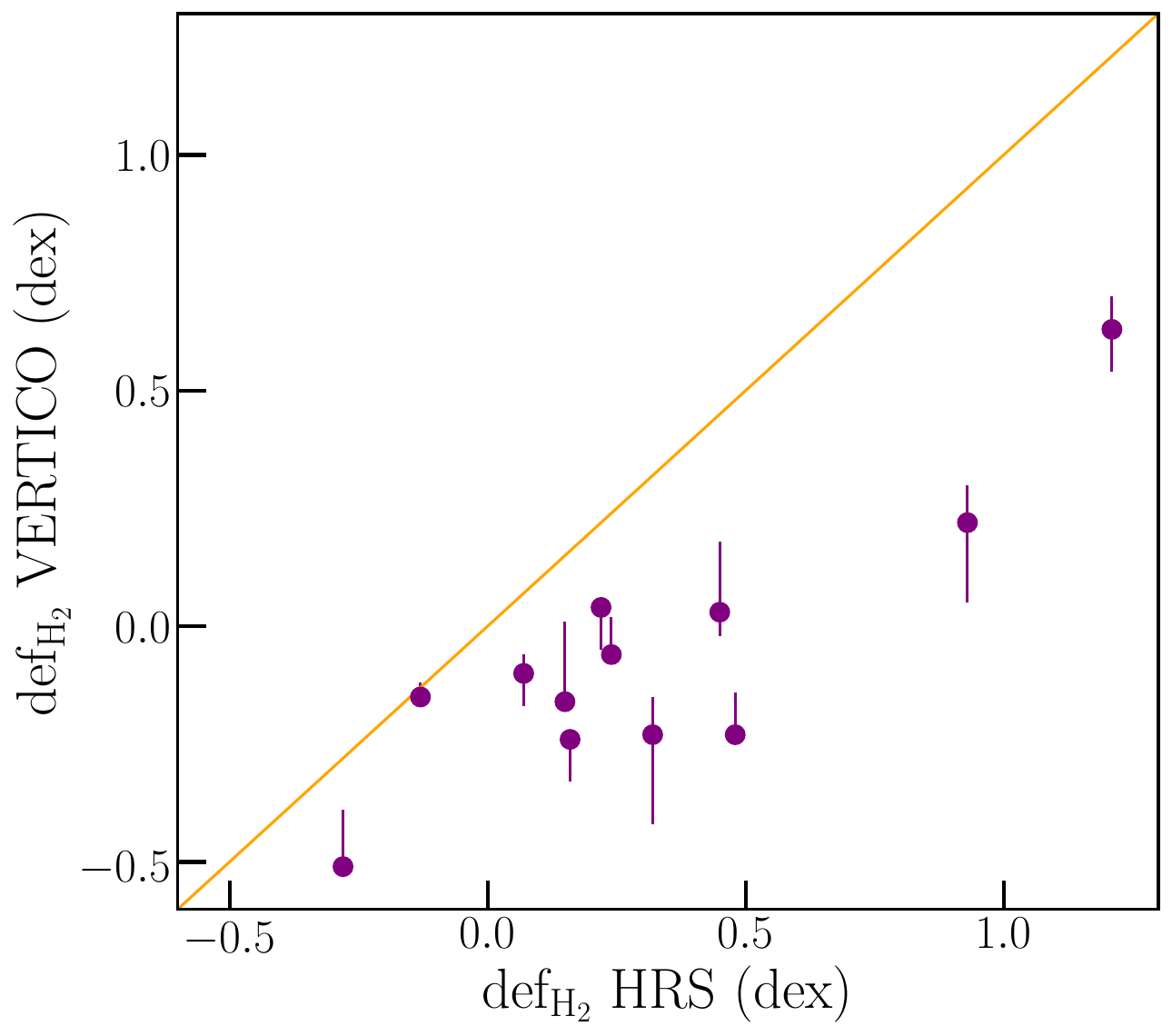}\label{fig:def_verti_hera}}
	
	\caption{\drafttwo{\emph{Left-hand panel:} \HI\ deficiencies of VERTICO galaxies from \citet{Chung2009}, based on $R_\star$, vs. \HI\ deficiencies based on the difference in \HI\ fraction compared to field galaxies at fixed stellar mass \draftfour{(see \S \ref{sub:HI_deficiency} for their respective definitions)}. The orange line represents the one-to-one relation. \HI\ deficiencies from \citet{Chung2009} \draftthree{correlate well with} those based on \HI\ fractions compared to the field. 
	\newline \emph{Right-hand panel:} H$_2$ deficiencies for a sub-sample of VERTICO that overlaps with high-quality data from the HRS. H$_2$ deficiencies for the VERTICO sample are systematically \draftthree{smaller} compared to those of the HRS, likely \draftfour{primarily} because of the difference in control samples used to calibrate this parameter.}}
	\label{fig:def_comparisons}
\end{figure}

\begin{sidewaysfigure}
	\centering
	\includegraphics[width=1\textwidth]{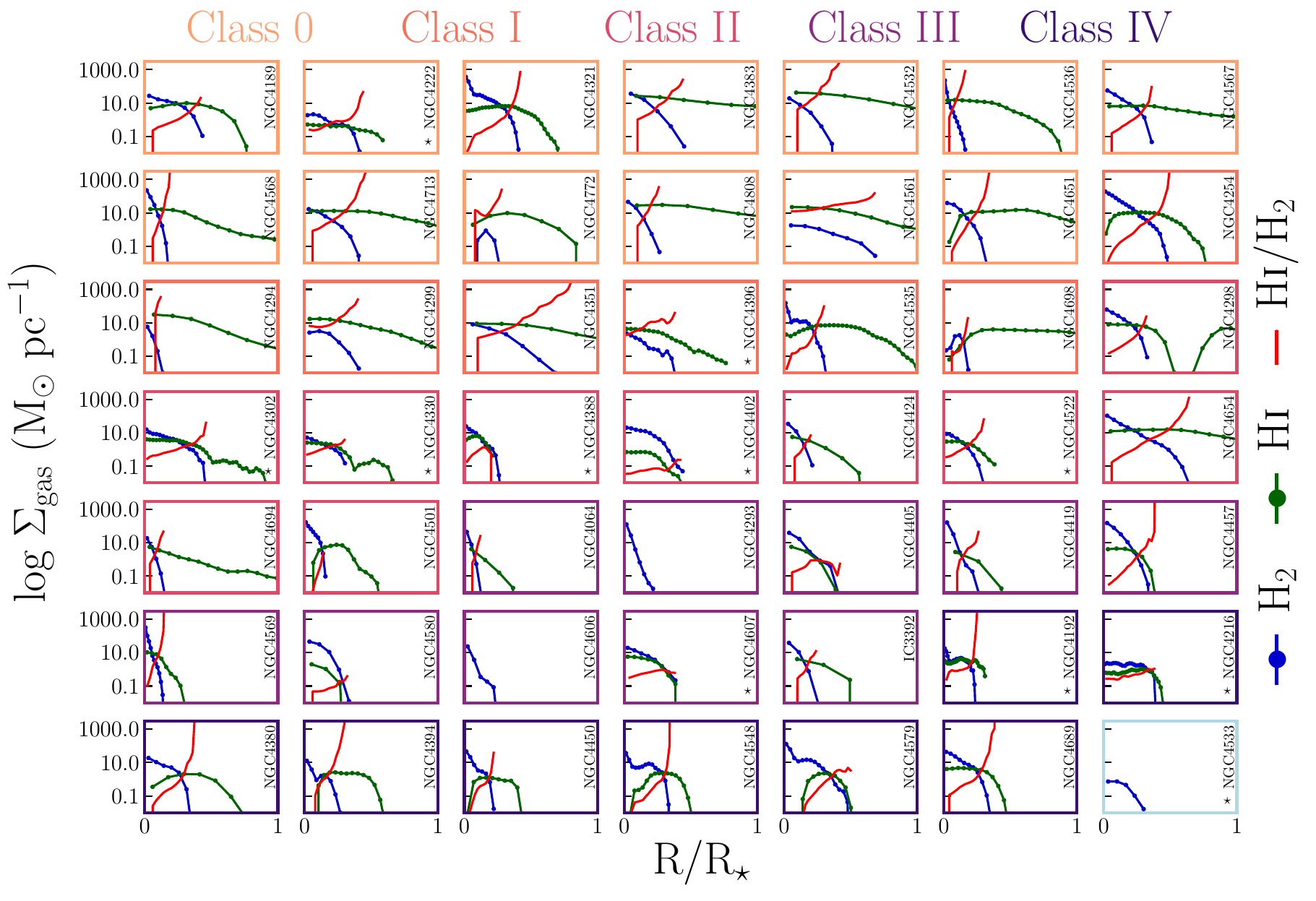}
	\caption{\draftthree{H$_2$ (blue) and \HI\ (green) radial profiles of all 49 CO-detected galaxies in VERTICO. The profile of the \HI-to-H$_2$ ratio is shown in red. The borders of the subplots are colour-coded by the VIVA classification of the galaxy (see \S \ref{sub:hi_classes}), indicated at the top of the Figure (no class is available for NGC4533, see \S \ref{sub:hi_classes}). Some galaxies (NGC4694, NGC4606, and NGC4293) only have marginal \HI\ detections. Therefore, no reliable radial profiles could be made for them. NGC4533 was not detected in \HI. Highly-inclined galaxies (inc $\geq$ 80$^o$) are indicated with a star in front of their name.}}
	\label{fig:rad_profs}
\end{sidewaysfigure}

\bibliography{../References.bib}{}
\bibliographystyle{aasjournal}



\end{document}